# A New Theoretic Foundation for Cross-Layer Optimization

Fangwen Fu* and Mihaela van der Schaar

## ABSTRACT

Cross-layer optimization solutions have been proposed in recent years to improve the performance of network users operating in a time-varying, error-prone wireless environment. However, these solutions often rely on ad-hoc optimization approaches, which ignore the different environmental dynamics experienced at various layers by a user and violate the layered network architecture of the protocol stack by requiring layers to provide access to their internal protocol parameters to other layers. This paper presents a new theoretic foundation for cross-layer optimization, which allows each layer to make autonomous decisions individually, while maximizing the utility of the wireless user by optimally determining what information needs to be exchanged among layers. Hence, this cross-layer framework does not change the current layered architecture. Specifically, because the wireless user interacts with the environment at various layers of the protocol stack, the cross-layer optimization problem is formulated as a layered Markov decision process (MDP) in which each layer adapts its own protocol parameters and exchanges information (messages) with other layers in order to cooperatively maximize the performance of the wireless user. The message exchange mechanism for determining the optimal cross-layer transmission strategies has been designed for both off-line optimization and on-line dynamic adaptation. We also show that many existing cross-layer optimization algorithms can be formulated as simplified, sub-optimal, versions of our layered MDP framework.

## I. INTRODUCTION

The Open Systems Interconnection (OSI) model [1] is a layered, abstract organization of the various communications and computer networks protocols. In the layered network architectures, the functionality

The authors are with the University of California Los Angeles, Dept. of Electrical Engineering (EE), 56-147 Engineering IV Building, 420 Westwood Plaza, Los Angeles, CA 90095; phone: 1-310-825-5843; fax: 1-310-206-4685; e-mail: fwfu@ee.ucla.edu (F. Fu); mihaela@ee.ucla.edu (M. van der Schaar).

*Corresponding author. Address and contact information: see above.

of each layer is specified in terms of the services that it receives from the layer(s) below it and that it is required to provide to the layer(s) above it. To optimize its service, each layer autonomously controls and optimizes a set of protocol parameters. However, the information exchange between multiple layers is limited in the current layered network architectures, which often results in sub-optimal performance of network users.

To optimize the different protocol parameters, the wireless stations (WSTAs) need to consider the dynamic wireless network "environment" resulting from the repeated interaction with other stations, the experienced time-varying channel conditions and, for delay-sensitive applications, the time-varying source characteristics. Moreover, it should be noted that a WSTA needs to jointly optimize the selected protocol parameters within each layer such that the utility of the WSTA is maximized. The joint optimization of the transmission strategies at the various layers is referred to as *cross-layer design* [2][3].

Recently, various cross-layer design methods have been proposed in order to jointly adapt the transmission strategies at each layer of the OSI stack to the rapidly varying environment and often scarce network resources. Current cross-layer design approaches can be coarsely categorized into two categories: *user-centric cross-layer optimization* and *network-centric cross-layer optimization*. These related works will be reviewed in Section II.

### *A. Remaining cross-layer design challenges*

The advantage of the layered architecture is that the designer or implementer of the protocol or algorithm at a particular layer can focus on that layer without worrying about the rest of stack [3]. However, most existing cross-layer design solutions advocate improving the system utility by violating the current layered architecture of wireless networks. These cross-layer interactions create the dependencies among the layers which will affect not only the concerned layer but also other layers. Hence, such solutions are undesirable because they require a complete redesign of current networks and protocols and thus, require a high implementation cost [3].

Furthermore, some existing cross-layer design solutions aim at maximizing the WSTA's utility by jointly adapting the transmission strategies across multiple layers to the current environmental dynamics [7][14][15]. These solutions, however, neglect that the environmental dynamics are also affected by the cross-layer transmission strategies, thereby affecting the future utility derived by the network users. In the literature, there are also several works solving the cross-layer design problem by considering the impact of current transmission strategies on the future reward [8][13][25][27][30][31] [33]. These works focus either on optimizing the decision process within a single layer (mostly the APP layer, e.g. [8][13]) or on selecting the joint action (i.e. the transmission strategy and parameters) across several layers, by assuming a centralized decision process, e.g. [25][27][30][31]. Hence, they do not consider the informationally-decentralized nature of the decision process, which is a byproduct of the current layered network architecture.

Therefore, in this paper, we focus on developing cross-layer solutions that *preserve* the current layered architecture of the protocol stack and that allow the layers to make decisions autonomously based on the dynamics they experience. Thus, the proposed cross-layer solutions are compliant with existing protocols and standards available at various layers (e.g. 802.11a [19], TCP, H.264 video coding etc.). We also focus on developing a new theoretic foundation for cross-layer optimization, which maximizes the utility obtained by WSTAs by optimally determining what information should be exchanged among layers and based on this information, what is the optimal decision that needs to be taken by each layer, in an autonomous manner.

### *B. Key features of the proposed framework*

Specifically, we propose a novel cross-layer design framework in which the transmission strategy is no longer decided based on myopic utility maximization. Instead, a foresighted transmission strategy is determined that explicitly considers the impact of the current action on the future utility. Similar to the works in [25][27][30][31], we model the cross-layer optimization problem as a Markov decision process (MDP) [21] that has as objective the maximization of the discounted sum of future utility. In this way, the impact of the currently selected cross-layer transmission strategy on the future utility (reward) is formulated in a systematic manner. This cross-layer design formulation will be detailed in Section IV.

Unlike the previous works that jointly optimize the cross-layer strategies in a centralized way, we propose a layered MDP solution to drive the cross-layer optimization, such that the resulting solution complies with the layered architecture and protocol separation implemented in current wireless networks. In this layered MDP framework, each layer makes its transmission decision (i.e. selects the transmission strategies, e.g. packet scheduling in the application (APP) layer, retransmission in the MAC layer and modulation selection in the physical (PHY) layer) in an autonomous manner, by considering the dynamics experienced at that layer as well as the information available from other layers. Importantly, using this layered optimization framework, we do not change the current layered architecture of the protocol stack. Moreover, the current algorithms and protocols currently implemented at each layer also remain unaffected, as the proposed framework requires only the exchange of information across layers and the optimization of available parameters at each layer. To exchange information across multiple layers, we define a message exchange mechanism in which the *content* of the message captures the performed transmission strategies and experienced dynamics at each layer. However, the *format* of the message is independent of the transmission strategies, protocols and dynamics implemented at each layer and can be implemented using any agreed-upon signaling protocol [29]. Hence, the various protocols can be kept the same, upgraded or entirely modified; the algorithms at the various layers can also be upgraded; and the supported applications can be changed without affecting the proposed cross-layer design framework. Furthermore, certain layers or algorithms can decide not to exchange any messages or to not participate in the cross-layer optimization.

In summary, this paper makes the following contributions:

- We propose a new theoretic cross-layer optimization framework which provides a systematic, rather than ad-hoc, mechanism for dynamically selecting and adapting the transmission strategy at each layer and the message exchange across layers. A layered MDP framework is proposed such that each layer makes its transmission decision autonomously, by considering its own experienced network dynamics. This layered optimization framework does not require a central decision maker to consider all the layers' parameters, constraints, protocols, algorithms etc. Hence, the layered architecture remains unaltered, thereby enabling a scalable, flexible and easily upgradable network design.
- A message exchange mechanism between the layers is developed, in which messages capture the experienced dynamics and the performed transmission strategies, but the format of the message is independent of the transmission strategies, protocols deployed and dynamics experienced at each layers. This proposed cross-layer optimization framework enables WSTA to easily upgrade the protocols and algorithms implemented at the various layers, and it does not require any changes to the current layered network architecture.
- The proposed layered MDP framework allows the designer to systematically simplify the cross-layer optimization by trading-off the performance and computation complexity. Furthermore, the layered MDP framework allows the designer to implement layered learning algorithms for on-line adaptation which adhere to the current layered network architecture.

### C. Paper Organization

The rest of the paper is organized as follows. Section II reviews the related literatures. Section III discusses the problem settings for the cross-layer optimization. Section IV formulates the cross-layer design as an MDP problem and briefly reviews the centralized value iteration algorithm. Section V presents a layered value iteration algorithm for optimally solving the layered MDP and discusses the simplifications of the layered MDP framework. Section VI discusses the on-line learning based on the layered MDP framework. Section VII gives an illustrative example for the layered MDP formulation of cross-layer optimization. The paper concludes in Section VIII.

## II. RELATED WORK

### A. User-centric cross-layer optimization

User-centric cross-layer optimization focuses on the transmission strategy adaptation at the user side [4]-[8]. The transmission strategies are mainly focused on adapting the per-packet transmission strategies, e.g. adaptive modulation and coding (AMC) in the PHY layer [16], automatic repeat request (ARQ) in the MAC layer [6] and priority scheduling in the APP layer [7]. The common assumption in this cross-layer adaptation is that the user experiences a stationary environment that can be characterized by a stationary stochastic process. For instance, in the PHY layer, the time-varying channel conditions – Signal to Noise Ratio (SNR) are characterized by independent transition [5][6] or Markovian transitions [4]. In the MAC layer, the channel access is characterized by opportunistic access [6] or static access [4][5]. A plethora of past research works [17] (and the references therein) model the channel as a Markov chain. These works only focus on the adaptation taking place at the one layer of the OSI stack, e.g. AMC in the PHY layer or priority scheduling in the APP layer.

Based on their objective, the user-level cross-layer optimization solutions can also be further classified into two main categories: quality-of-service (QoS)-oriented cross-layer adaptation [4][5][6] and utility-oriented cross-layer adaptation [7] [8]. The QoS-oriented cross-layer adaptation aims at improving the QoS for the supported applications, such as the guaranteed delay and goodput. In this type of cross-layer adaptation, the data is often assumed to arrive according to a well-known distribution such as a Poisson distribution and, given this distribution, the average delay and goodput are determined. Alternatively, the utility-oriented cross-layer adaptation *explicitly* considers the characteristics of the applications such as the packet-based delay constraints, packet priorities as well as packet dependencies, which are often time-varying due to the dynamically changing source characteristics and selected encoding parameters [8]. Another key difference between these two cross-layer adaptation solutions is that the former considers only a coarse prioritization of the traffic based on the resulting utility impact, for instance by adopting priority queuing, while the later can perform fine-granularity optimization by explicitly considering the various packets' delay deadline, utility impact, etc.

### B. *Network-centric cross-layer optimization*

Network-centric cross-layer optimization considers a multi-user network and aims at providing efficient joint adaptation across the various layers of the OSI stack in order to improve the network utility, rather than the user utility (in contrast to the user-centric optimization). For example, the network utility can be defined as the sum of all users' utilities. Hence, the network-centric cross-layer optimization is generally formulated as a network utility maximization (NUM). A comprehensive overview of NUM can be found in [9] and the references therein. In the NUM framework, the cross-layer design has been formulated as an optimization problem by maximizing the sum of network users' utilities given the resource constraints [10]-[13]. The NUM problem is generally decomposed into several sub-problems and solved via distributed algorithms.

The solutions to the NUM problem assume that (i) the network users' objective can be perfectly characterized by a static utility function; (ii) each user possesses the accurate aggregated information about the network congestion; (iii) each user has the capability to continuously and smoothly adapt its transmission strategies according to the differentiable utility function. However, the NUM framework has several key limitations for cross-layer optimization in wireless networks. First and most importantly, the NUM framework does not consider the different dynamic information at the various layers of the OSI stack and the granularity of decisions at each layer. Second, the changing application and source characteristics often lead to time-varying and non-differentiable utility functions [8], and hence, the convergence of NUM is not guaranteed such that this optimization often leads to suboptimal solutions. Third, the aggregated information may generally suffer from random errors due to the error-prone measurement and transmission. Forth, the decision granularity for each WSTA should depend on its available transmission strategies, algorithms, protocols or computational resources. All these challenges are ignored by the NUM formulation, and hence, this cannot provide suitable cross-layer solutions aimed at optimizing the individual utility of autonomous network users.

# III. CROSS-LAYER PROBLEM STATEMENT

We consider one WSTA transmitting its time-varying traffic to another WSTA (e.g. base station) over a wireless network (e.g. wireless LAN, cellular network, etc.). We also assume that there are $L$ participating layers[1] in the protocol stack. Each layer is indexed $l \in \{1,...,L\}$ with layer 1 corresponding to the lowest participating layer (e.g. PHY layer) and layer $L$ corresponding to the highest participating layer (e.g. APP layer). In this paper, we focus on the cross-layer adaptation of the $L$ layers of a WSTA, i.e. user-centric cross-layer adaptation. As shown in Figure 1, the WSTA interacts with the dynamic environment at various layers in order to maximize the application utility.

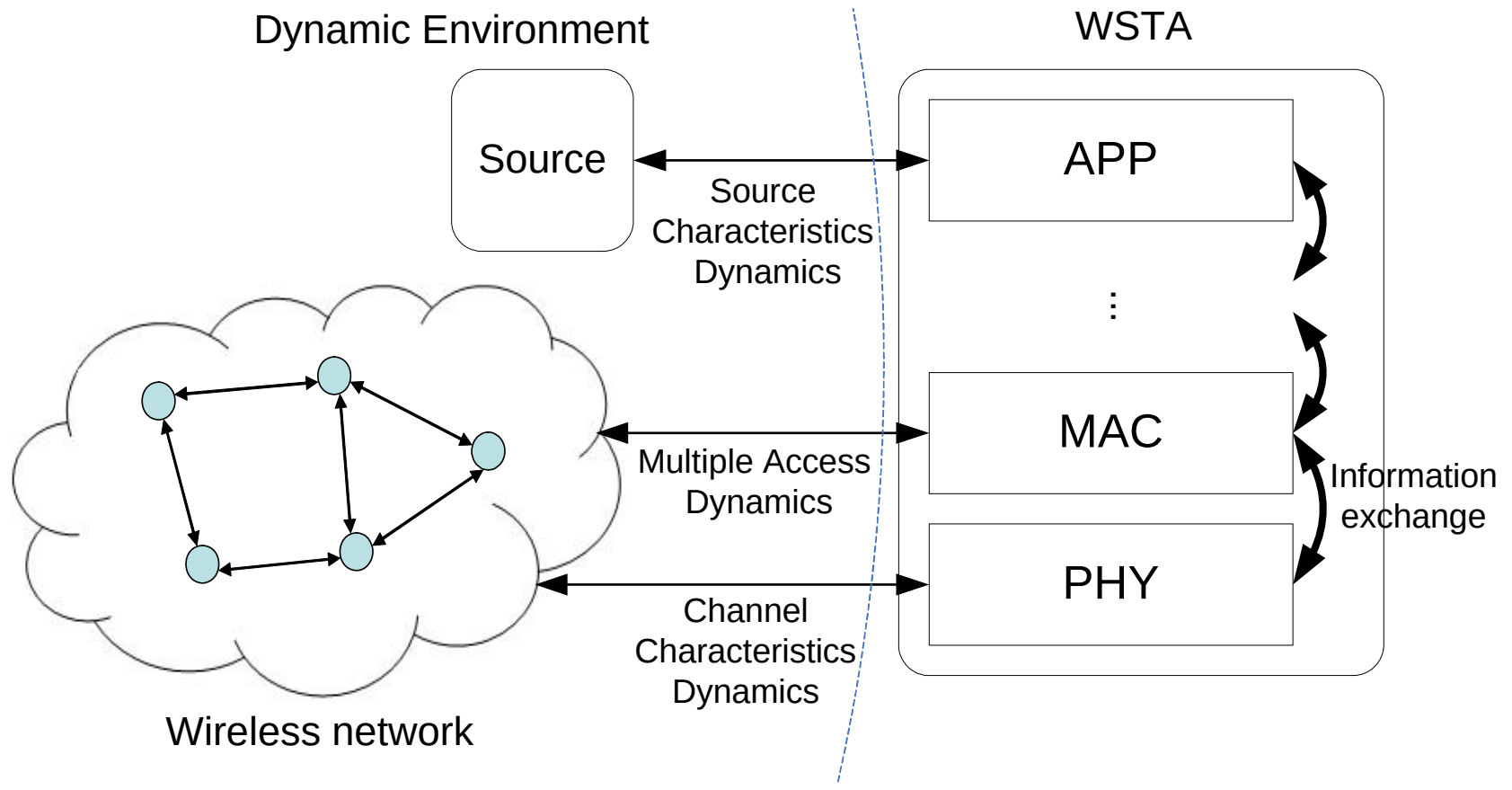

Figure 1.WSTA interacts with the dynamic environment at multiple layers of the OSI stack.[2]

*Example 1*: Although the cross-layer optimization framework proposed in this paper is general, we would like to first provide a concrete example of a cross-layer optimization problem, in order to help the readers become familiar with the concept of actions and states defined in Section III.A-B. Similar to [25], in this example, we consider that the WSTA transmits delay-sensitive data to another WSTA and accesses the wireless channel using TDMA (e.g. as in 802.11e HCF [20]). Assume that the time is slotted and divided into frames consisting of $N$ time slots. We consider the optimization of the transmission strategies available at the APP, MAC, and PHY layers, i.e. $L = 3$.

In the PHY layer, the channel gain can be modeled as a finite state Markov chain (FSMC) [26]. To satisfy the service requirement from upper layers, the PHY layer adapt its transmission power level, and the modulation schemes based on the channel dynamics.

[1] If one layer does not participate in the cross-layer design, it can simply be omitted. Hence, we consider here only the $L$ participating layers.
[2] Generally, the information exchange can be performed between arbitrary layers.

In the MAC layer, the number of time slots allocated to the WSTA depends on the scheduling algorithm deployed in the network, e.g. round-robin scheduling, max-gain scheduling [32]. However, in this paper, we consider a more general multi-user scheduling method which allows the WSTA to dynamically compete for the available time slots [2]. At the beginning of each frame, the WSTA competes with other WSTAs for the time to access the spectrum. We can model the amount of time allocated to the WSTA as a finite state Markov chain, which is controlled by the competition bid. If the round-robin or max-gain scheduling algorithms are used, the competition actions are empty. Besides competing for the resource, the MAC can also perform error control algorithms (e.g. ARQ) to improve the service provided to the upper layers.

In the APP layer, the WSTA generates delay-sensitive data. The delay-sensitivity is represented by the delay deadlines after which the packets will expire and therefore not contribute to the WSTA's utility. As in [25], we can model the number of packets with the various delay deadlines available for transmission as a Markov chain. The number of packets available for transmission depends on the source coding parameters adaptation as well as the transmission strategies at the lower layers.

The objective of the WSTA is to jointly adapt the transmission strategies across all the three layers such that the system utility is maximized. The simulation results for this example will be given in Section VII. ■

*A. States*

In this paper, the state of the layers is defined such that future transmission strategies can be determined independent of the past history given the current state. In other words, the state encapsulates all the past information required for future strategy adaptation. We refer to this type of state as Markovian. When considering the layered architecture of current networks, we are able to define a state $s_l \in \mathcal{S}_l$ for each layer $l$. Then, the state of the entire WSTA is denoted by $\boldsymbol{s} \in \boldsymbol{\mathcal{S}}$, with $\boldsymbol{\mathcal{S}} = \prod_{l=1}^{L} \mathcal{S}_l$.

*Example 2*: The states for each layer in example 1 are defined as follows. At the PHY layer, the channel gain can be modeled as a FSMC. Hence, we can define the state of the PHY layer as the channel gain.The state of the MAC layer can be the number of time slots allocated to the WSTA in the current frame. As shown in [25][32], the number of time slots can be modeled as a FSMC. At the APP layer, the state can be defined as the number of packets having different delay deadlines. As shown in [25], the states at the APP layer also follow the FSMC model. ■

We note that states can be similarly defined for other layers, and that they can be more sophisticated than the simple example here.

*B. Actions*

In a layered architecture, a WSTA takes different transmission actions in each state of each layer. The transmission actions can be classified into two types at each layer $l$: an external action is performed to determine the state transition, and an internal action is performed to determine the service provided to the upper layers for the packet(s) transmission.

The external actions at each layer $l$ are denoted by $A_l \in \mathcal{A}_l$, where $\mathcal{A}_l$ is the set of the possible external actions available at layer $l$. The external actions for the WSTA in all the layers are denoted by $\boldsymbol{A} = [A_1,...,A_L] \in \boldsymbol{\mathcal{A}}$, where $\boldsymbol{\mathcal{A}} = \prod_{l=1}^{L} \mathcal{A}_l$. The internal actions are denoted by $B_l \in \mathcal{B}_l$, where $\mathcal{B}_l$ is the set of the possible internal actions available at layer $l$. The internal actions are performed by the WSTA to efficiently *utilize* the wireless medium given the network resource allocation and its own resource budget (e.g. power constraint), by providing the QoS required by the supported applications. The internal actions for the WSTA across all the layers are denoted by $\boldsymbol{B} = [B_1,...,B_L] \in \boldsymbol{\mathcal{B}}$, where $\boldsymbol{\mathcal{B}} = \prod_{l=1}^{L} \mathcal{B}_l$. Hence, the action at layer $l$ is the aggregation of external and internal actions, denoted by $\Psi_l = \left[ A_l \ B_l \right] \in \boldsymbol{\mathcal{X}}_l$, where $\boldsymbol{\mathcal{X}}_l = \mathcal{A}_l \times \mathcal{B}_l$. The joint action of the WSTA is denoted by $\boldsymbol{\Psi} = [\Psi_1,...,\Psi_L] \in \prod_{l=1}^{L} \boldsymbol{\mathcal{X}}_l$. The external and internal actions performed in the WSTA are illustrated in Figure 2.

*Example 3*: In example 1, the external actions performed to determine the state transition can be the competition bids for the resource allocations in the MAC layer [6], and the adaptation of the source coding parameters in the APP layer. The external actions at the PHY layer are empty because the channel gain transition is fully determined by the environmental dynamics. The internal actions at the PHY layer can be the power allocation and modulation schemes. The internal actions in the MAC layer can be the adaptation of the retransmission limit. The internal actions at the APP layer are not taken into account and hence, are empty. ■

In this paper, we further allow the WSTA to perform mixed actions. The mixed action is defined as a vector of probabilities that the WSTA assigns to its selecting action. In our context, the external mixed action at layer $l$ is denoted by $a_l \in \Delta_l^a$, the internal action at layer $l$ is denoted by $b_l \in \Delta_l^b$, where $\Delta_l^a$ and $\Delta_l^b$ is the set of the mixed actions for external and internal actions, respectively. The mixed action at layer $l$ is denoted by $\xi_l = [a_l, b_l] \in \Delta_l^a \times \Delta_l^b$. The joint mixed action of the WSTA is denoted by $\boldsymbol{\xi} = [\xi_1,...,\xi_L] \in \prod_{l=1}^{L} \Delta_l^a \times \Delta_l^b$.

By splitting the transmission actions into the internal and external actions, we have the following advantages, which will become clear in Section V:

- The current utility computation based on the internal actions is separated from the state transition based on the external actions at each layer. This separation enables us to design a layered cross-layer optimization framework.
- The separation between the internal actions and external actions enables us to design an interlayer message exchange mechanism that is independent of the specific format of the protocols and algorithms deployed at each layer.

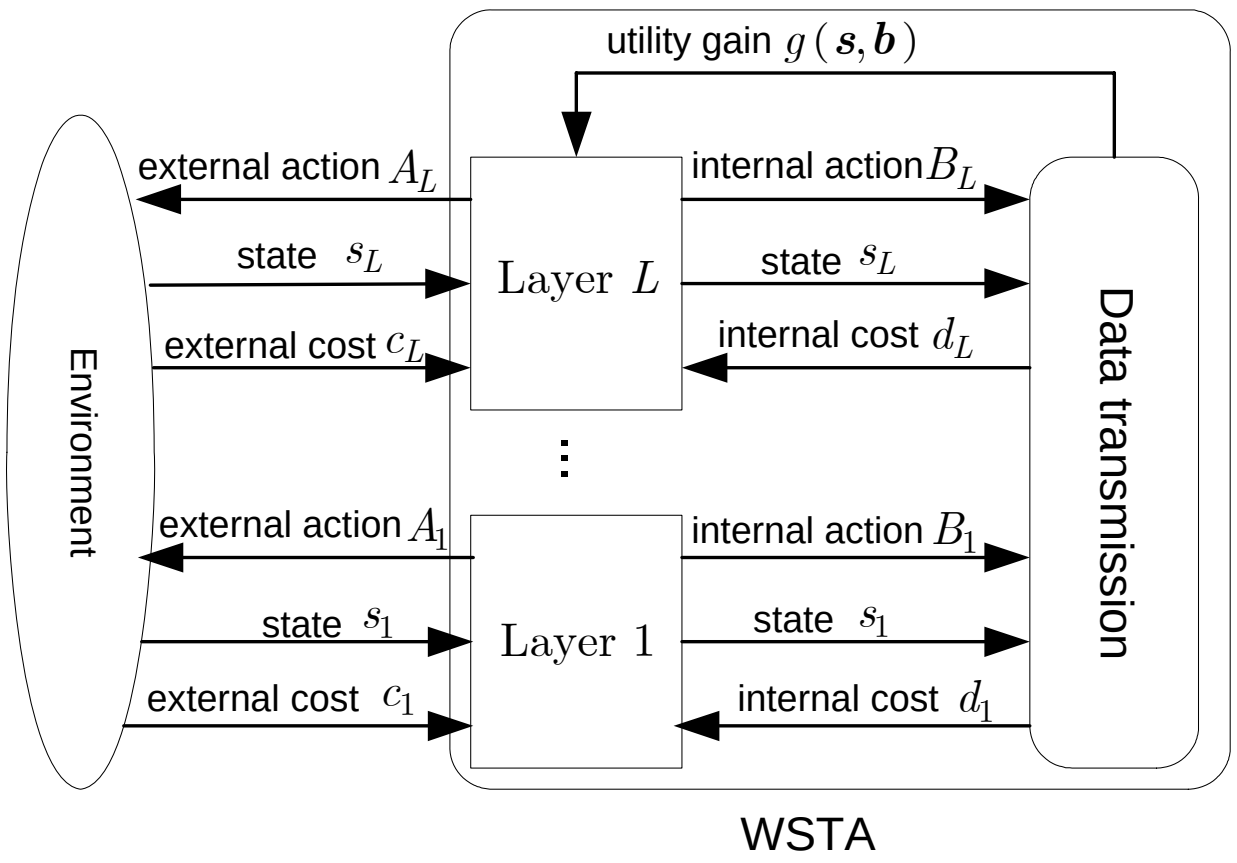

Figure 2.Illustration for the states, internal and external actions and cost at each layer, and transmission gain at layer $L$

*C. Transition probability*

In Section III.B, the external actions are performed to drive the state transition. In this section, we examine the transition probability and the underlying models for environmental dynamics. In general, because states are Markovian, the state transition of the WSTA only depends on the current state $\boldsymbol{s}$, the current performed actions, and the environmental dynamics. The corresponding transition probability is denoted by $p(\boldsymbol{s}' \mid \boldsymbol{s}, \boldsymbol{\xi})$.

Due to the layered architecture of the wireless network, the state transition probability can be further decomposed. Using Bayes rule, the transition probability can be rewritten as

$$p(\boldsymbol{s}' \mid \boldsymbol{s}, \boldsymbol{\xi}) = \prod_{l=1}^{L} p(s_l' \mid \boldsymbol{s}_{1\to l\text{-}1}', \boldsymbol{s}, \boldsymbol{\xi}) \tag{1}$$

where $\boldsymbol{s}_{1\to l}' = [s', ..., s_l']$.

*Example 4*: In example 1, the state transition in the APP layer depends on the states and internal actions at all the layers (since they will affect the amount of packets to be transmitted) and the external action at the APP layer. Hence, the transition probability is given by $p\left(s_3' \mid \boldsymbol{s}, a_3, \boldsymbol{b}\right)$. (Recall that $s_1, s_2$ and $s_3$ are the states of the PHY, MAC and APP layers, respectively.) The transition probability at the PHY layer depends only on the current state. Hence, the transition probability is given by $p\left(s_1' \mid s_1\right)$. At the MAC layer, the state transition depends on the deployed scheduling algorithm. For example, when the round-robin scheduling is used, the state transition probability becomes $p\left(s_2' \mid s_2\right)$ since the external action is empty [32]. When the max-gain scheduling is used, the state transition probability becomes $p\left(s_2' \mid s_1', s_2\right)$ since the state of the PHY layer is required for the network coordinator to schedule the network users [32]. When the WSTA competes for the network resources at the MAC layer, the state transition probability becomes $p\left(s_2' \mid s_2, a_2\right)$. ■

In this paper, based on the actions we illustrate in Section III.B, the transition probability can be decomposed as

$$p\left(\boldsymbol{s}' \mid \boldsymbol{s}, \boldsymbol{\xi}\right) = \prod_{l=1}^{L-1} p\left(s_l' \mid \boldsymbol{s}_{1\to l\text{-}1}', s_l, a_l\right) p\left(s_L' \mid \boldsymbol{s}_{1\to L\text{-}1}', \boldsymbol{s}, a_L, \boldsymbol{b}\right). \tag{2}$$

Note that the transition probability given in example 3 satisfies Eq.(2).

This decomposition is due to the layered network architecture and enables us to develop a layered MDP framework, which will be presented in Section V.

*D. Utility function*

The utility gain obtained in layer $L$ is based on the states and internal actions at each layer and it is denoted by $g(\boldsymbol{s}, \boldsymbol{b})$. The transmission cost at layer $l$ represents the cost of performing both the external and internal actions, e.g. the amount of power allocated to determine the channel conditions or the tax (tokens, money) spent for consuming wireless resources [23][24]. In general, the transmission cost of performing the external (internal) action at layer $l$ is denoted by $c_l(s_l, a_l)$ ($d_l(s_l, b_l)$), which is a function of the external (internal) action and the state of layer $l$. The utility gain and layer costs are depicted in Figure 2. For illustration, we assume that the reward is defined as

$$R(\boldsymbol{s}, \boldsymbol{\xi}) = g(\boldsymbol{s}, \boldsymbol{b}) - \sum_{l=1}^{L} \lambda_l^a c_l(s_l, a_l) - \sum_{l=1}^{L} \lambda_l^b d_l(s_l, b_l) \tag{3}$$

where $\lambda_l^a$ ($\lambda_l^b$) is a external (internal) Lagrangian multiplier in layer $l$, determined by the WSTA to trade off the utility and transmission cost. We assume that the Lagrangian multipliers $\lambda_l^a$ and $\lambda_l^b$ are known. The optimal Lagrangian multipliers depend on the available resource budget and can be obtained as in [27]. The reward in Eq. (3) can be further decomposed into two parts: one is the internal reward, which depends on the internal actions, and the other is the external reward, which depends on the external actions. The internal reward is

$$R_{in}(\boldsymbol{s},\boldsymbol{b}) = g(\boldsymbol{s},\boldsymbol{b}) - \sum_{l=1}^{L} \lambda_l^b d_l(s_l, b_l), \tag{4}$$

and the external reward is

$$R_{ex}(\boldsymbol{s},\boldsymbol{a}) = -\sum_{l=1}^{L} \lambda_l^a c_l(s_l, a_l). \tag{5}$$

Hence, the reward is $R = R_{in} + R_{ex}$.

*Example 5*: In example 1, the utility function can be defined as

$$R(\boldsymbol{s},\boldsymbol{\xi}) = g(\boldsymbol{s},\boldsymbol{b}) - \lambda_2^a c_2(s_2, a_2) - \lambda_1^b d_1(s_1, b_1) \tag{6}$$

where $c_2(s_2, a_2)$ represents the competition bids spent for resource negotiation in the MAC layer by performing the external action $a_2$ and $d_1(s_1, b_1)$ represents the cost of the allocated power at the PHY layer which depends on the power allocation in the internal action. The internal reward is $R_{in}(\boldsymbol{s},\boldsymbol{b}) = g(\boldsymbol{s},\boldsymbol{b}) - \lambda_1^b d_1(s_1, b_1)$ and the external reward is $R_{ex}(\boldsymbol{s},\boldsymbol{a}) = -\lambda_2^a c_2(s_2, a_2)$. ■

# IV. Foresighted Cross-Layer Decision Framework

## A. Foresighted decision making

As described in Section III.B, the state transition at each layer is controlled by the external actions. For simplicity, we assume that the state transition in each layer is synchronized and operates at the same time scale, such that the transition can be discretized into stages during which the WSTA has constant state and performs static actions. The length of the stage can be determined based on how fast the environment changes (e.g. the stage in the example 1 corresponds to one frame with $N$ time slots). We use a superscript $k$ to denote stage $k$. Hence, the state of the WSTA at stage $k \in \mathbb{N}$ is denoted by $\boldsymbol{s}^k$ with each element $s_l^k$ being the state of layer $l$; similarly, the joint action performed by the WSTA at state $k$ is $\boldsymbol{\xi}^k$ with each element $\xi_l^k = [a_l^k, b_l^k]$. The state transition probability is given by Eq. (2) and the stage reward is given by Eq. (3).

Unlike the tradition cross-layer adaptation that focuses on the myopic (i.e. immediate) utility, in the proposed the cross-layer framework, the goal is to find the optimal internal and external actions at each stage such that a cumulative function of the rewards is maximized. We refer to this decision process as the *foresighted* cross-layer decision. By maximizing the cumulative reward, the WSTA is able to take into account the impact of the current actions on the future reward.

Specifically, we assume that the WSTA will maximize the discounted accumulative reward, which is defined as

$$\sum_{k=0}^{\infty}(\gamma)^k R\left(\boldsymbol{s}^k, \boldsymbol{\xi}^k \mid \boldsymbol{s}^0\right) \tag{7}$$

where $\gamma$ is a discounted rate with $0 \leq \gamma < 1$ and $\boldsymbol{s}^0$ is the initial state. Unlike the formulation in [27][33], where the time-average reward is considered, we use the discounted accumulated reward with higher weight on the current reward. The reasons for this are as follows: (i) for delay-sensitive applications, the data needs to be sent out as soon as possible to avoid expiration, and (ii) due to the unexpected environmental dynamics in the future, the WSTA may care more about the immediate reward. Hence, this needs to be considered when determining the values of $\gamma$ for a specific cross-layer problem

*B. Centralized cross-layer optimization*

As discussed in Section I, the internal and external actions need to be jointly optimized in order to determine the optimal cross-layer performance. Hence, information exchanges between layers are required. Existing cross-layer optimization frameworks require a central controller to decide the parameter configuration assuming that the complete information from all the layers is available to the central controller [4]-[8]. The foresighted cross-layer optimization can be formulated as an MDP which is defined as follows:

**Definition 1.** (**MDP**) An MDP is defined [21] as a tuple $M = \langle \boldsymbol{\mathcal{S}}, \boldsymbol{\mathcal{X}}, p, R, \gamma \rangle$ where $\boldsymbol{\mathcal{S}}$ is a joint state space, i.e. , $\boldsymbol{\mathcal{X}}$ is a joint action space for each state, $p$ is a transition probability function $\boldsymbol{\mathcal{S}} \times \boldsymbol{\mathcal{X}} \times \boldsymbol{\mathcal{S}} \mapsto [0,1]$, $R$ is a reward function $\boldsymbol{\mathcal{S}} \times \boldsymbol{\mathcal{X}} \mapsto \Re$ and $\gamma$ is the discounted factor.

In our context, the joint state space is $\boldsymbol{\mathcal{S}} = \prod_{l=1}^{L} \mathcal{S}_l$, the joint action space is given by $\boldsymbol{\mathcal{X}} = \prod_{l=1}^{L} \boldsymbol{\mathcal{X}}_l$, the transition probability is given by Eq. (2) and the reward function is given by Eq (3).

Similar to [13][25][27], the foresighted cross-layer optimization can be solved in a centralized way. Based on this complete information, the central optimizer is able to find the optimal decision rules for determining the internal and external actions at each layer. To solve the MDP problem, the central optimizer needs to know the following (see Figure 3 (a)):

- the state space at each layer;
- the action space at each layer;
- probability distribution describing the state transition (i.e. environmental dynamics);
- state reward function of the states and performed actions;

Several centralized algorithms (e.g. the policy iteration, value iteration and linear programming [22]) have been proposed to find the optimal policy which maximizes the discounted sum of future reward. However, most of the algorithms neglect the layered structure of the cross-layer optimization. In this section, we briefly review the value iteration algorithm which serves as a guideline for deriving the layered MDP model with information exchange in Section V.

*1) Policy*

In the MDP problem, the actions are selected based upon the past history, $\mathcal{H}$, that the WSTA experiences. The policy $\pi$ is referred to as a mapping from the experienced history $\mathcal{H}$ to the possible action set, i.e. $\pi : \mathcal{H} \mapsto \prod_{l=1}^{L} \mathcal{X}_l$. The history $\mathcal{H}$ includes the starting state, subsequent states and the actions taken up to the current stage, i.e. $\mathcal{H} = \left\{ \boldsymbol{s}^0, \boldsymbol{\xi}^0, \boldsymbol{s}^1, \boldsymbol{\xi}^1, ..., \boldsymbol{s}^k \right\}$. It has been shown that when the state of the MDP is fully observable, the optimal utility can be achieved using only the current state to decide what action to take [22]. A policy that uses only the current state is called a *Markov* policy. Since we assume that each layer is able to observe its own state by taking the external action, implementing a Markov policy for the WSTA requires layers to exchange their state information. For the Markov policies, we can further define a *decision rule* which is a mapping from the set of states to the set of actions at each stage $k$, i.e. $\phi^k : \boldsymbol{\mathcal{S}} \mapsto \prod_{l=1}^{L} \Delta_l^a \times \Delta_l^b$. For an infinite horizon MDP, a Markov policy $\pi$ contains a sequence of decision rules, i.e. $\pi = \left[ \phi^0, \cdots, \phi^k, \cdots \right]$. A policy $\pi$ is called a *non-stationary* policy if its decision rule at each stage is different. On the other hand, a policy $\pi$ is called a *stationary* policy if all the decision rules are the same. In general, a stationary policy can be found for an infinite horizon MDP [22]. Hence, we only consider the stationary and Markov policy in this paper, i.e. $\pi = \phi = \left[ \phi^a, \phi^b \right]$.

*2) State-value function and value iteration*

A state-value function is used to evaluate the performance of a cross-layer policy $\pi$ at each state. Based on the discounted sum of future reward function defined in Eq. (7), we can compute the state-value function for the policy $\pi$ in a recursive form:

$$V(\pi, \boldsymbol{s}) = \underbrace{R(\boldsymbol{s}, \pi(\boldsymbol{s}))}_{\substack{\text{stage reward} \\ \text{at current time}}} + \underbrace{\gamma \sum_{\boldsymbol{s}' \in \boldsymbol{\mathcal{S}}} p\left(\boldsymbol{s}' \mid \boldsymbol{s}, \pi(\boldsymbol{s})\right) V\left(\pi, \boldsymbol{s}'\right)}_{\text{expected future reward}}. \quad (8)$$

The state-value function in Eq. (8) includes two parts: one is the immediate reward and the other one is the discounted expected future reward. From Eq. (8), the state-value function of the policy $\pi$ can be computed by solving a linear program for any state $\boldsymbol{s}$ [22].

Using the state-value function, we can compare two different policies as follows: for different policies $\pi$ and $\pi'$, if $V(\pi,\boldsymbol{s}) > V(\pi',\boldsymbol{s})$, then we say the policy $\pi$ is better than $\pi'$ when the process starts from state $\boldsymbol{s}$.

In the proposed cross-layer foresighted decision problem, we aim at finding the optimal policy that maximizes the discounted accumulated rewards given in Eq. (7). By taking advantage of the principle of optimality of an MDP [22] and using dynamic programming, the optimal value function $V^*$ and corresponding optimal policy $\pi^*$ can be iteratively computed as follows:

$$V_n^*(\boldsymbol{s}) = \max_{\boldsymbol{\xi}\in\prod_{l=1}^{L}\Delta_l^a\times\Delta_l^b}\left\{R(\boldsymbol{s},\boldsymbol{\xi}) + \gamma\sum_{\boldsymbol{s}'\in\boldsymbol{\mathcal{S}}} p(\boldsymbol{s}' \mid \boldsymbol{s},\boldsymbol{\xi}) V_{n-1}^*(\boldsymbol{s}')\right\}. \quad (9)$$

where $n$ is the index of the iteration. The optimal stationary policy $\pi^*$ is obtained when $n \to \infty$.

The computation for the optimal value function and corresponding optimal policy shown in Eqs. (9) is called value iteration. The value iteration procedure can be found in [22]. where it has been proven that this procedure will converge to the optimal value function and optimal policy for the infinite horizon MDP. The value iteration procedure for finding optimal internal and external actions of the MDP will play a central role in the layered MDP model with information exchange for the foresighted cross-layer optimization.

### C. *Limitations associated with centralized cross-layer optimization*

In the centralized optimization described in Section IV.B, the actions at all the layers are selected simultaneously by the cross-layer optimizer. However, this centralized optimization exhibits the following problems when implemented in the layered network architectures.

First, from Figure 3 (a), it is clear that the centralized cross-layer optimization solution requires each layer to forward the complete information about its protocol-dependent dynamics, as well as its internal and external action space and state space to the central optimizer. This centralized decision violates the current layered network architecture [3]. Specifically, a completely new interface between the central optimizer and all the layers is created. The central optimizer is allowed to access the internal variables at each layer and hence, it is required to know the details about the protocols and algorithms deployed at each layer.

Second, the centralized optimization obliges each layer to take actions specified by the central optimizer.

The layers have no freedom to adapt their own actions to the environmental dynamics which they experience. Hence, inherently, each layer loses the power to design its own protocol independently of other layers, which inhibit the upgrade of layers' protocols and algorithms.

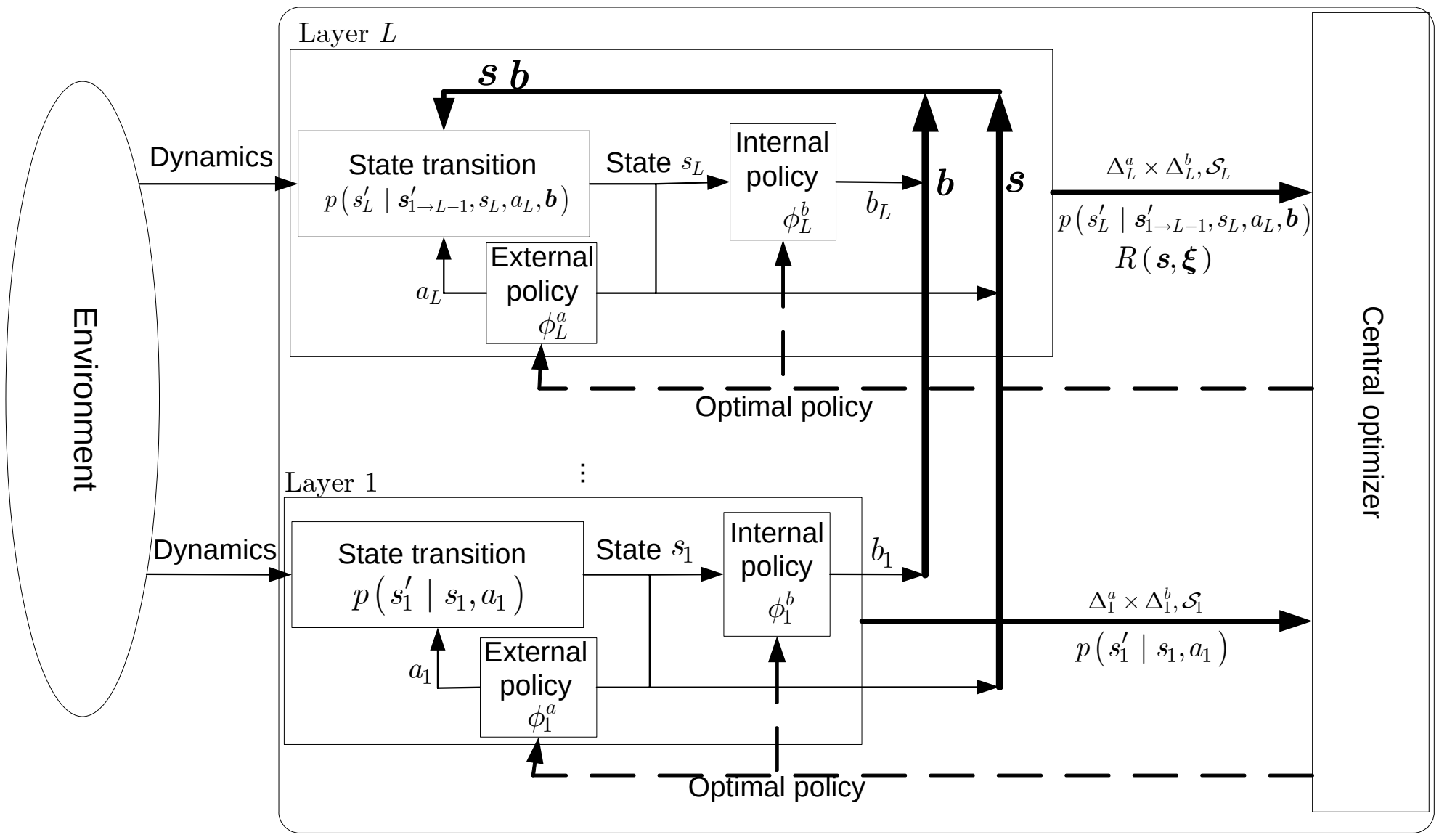

(a) Centralized cross-layer optimization framework

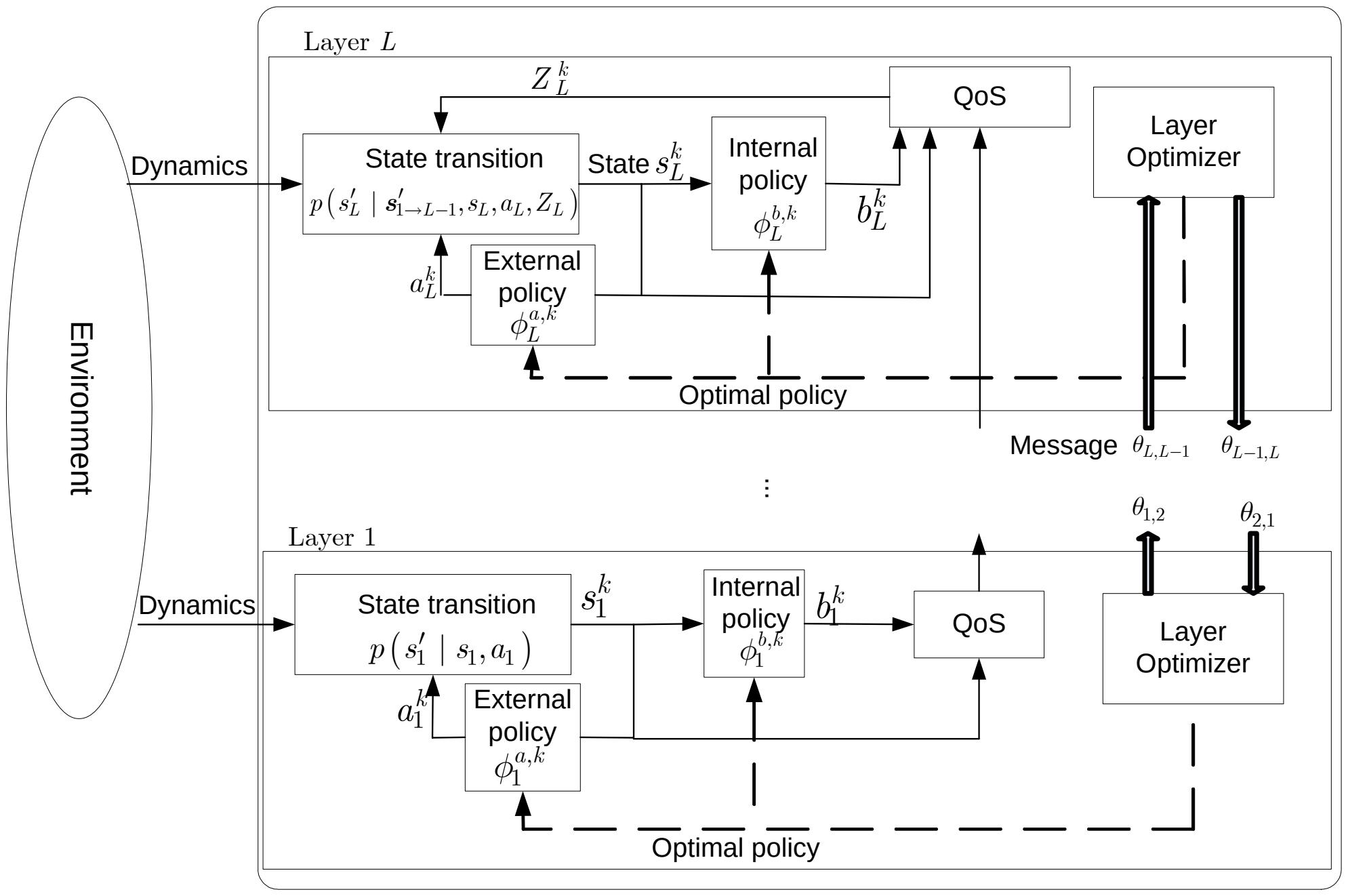

(b) Layered cross-layer optimization framework

Figure 3. Comparison of traditional cross-layer optimization framework and proposed cross-layer optimization framework

## V. LAYERED VALUE ITERATION FOR CROSS-LAYER FORESIGHTED DECISION

Although the centralized foresighted cross-layer optimization (formulated as the MDP problem) considers the information exchange among the layers in an indirect way, it unfortunately violates the layered architecture.

To overcome the problems associated with the centralized cross-layer optimization, in this paper we propose a layered foresighted cross-layer decision framework, which enables the layers to make optimal decisions on the transmission actions autonomously and to exchange information. In this way, the layered architecture is kept unchanged.

### *A. Layered MDP with information exchange*

Generally speaking, the cross-layer optimization allows each layer to communicate with any other layer. To adhere to the current layered architecture, we consider that the information is exchanged between the neighboring layers. Such information exchange does not result in any performance loss in our framework compared to the full information exchange between all layers for the centralized cross-layer optimization as shown in Figure 3(a) or as in [18]. This will be proved in Section V.C. Communication between neighboring layers requires only minimal changes to the current layered architecture. The message sent by layer $l$ to layer $l'$ at stage $k$ is denoted by $\theta_{ll'}^k$. Based on the above restriction, it is clear that $\theta_{ll'}^k = \varnothing$ , if $l' \notin \{l-1, l+1\}$. The message $\theta_{ll'}^k$ from the lower layer to the higher layer (i.e. $l < l'$) is called *upward* message and the message from the higher layer to the lower layer (i.e. $l > l'$) is called *downward* message.

The details of the message will be discussed in Sections V.B and V.C. Using this message exchange, each layer can decide its transmission strategy autonomously, based only on its own state and the information exchanged with the adjacent layers.

**Definition 2.** (**Layered MDP with information exchange**) The layered MDP model with information exchange is given by the tuple $\mathcal{M} = \left\langle \mathcal{L}, \mathcal{S}, \{\mathcal{X}_l\}_{l=1}^L, \{\Theta_{l,l+1}\}_{l=1}^{L-1}, \{\Theta_{l,l-1}\}_{l=2}^{L}, p, R, \gamma \right\rangle$, where

- $\mathcal{L} = \{1, ..., L\}$ is a set of $L$ layers, each of which takes the internal and external actions individually.
- $\mathcal{S}$ is a finite set of states, each element $\boldsymbol{s} \in \mathcal{S}$ of which contains $[s_1, \cdots, s_L]$.
- $\mathcal{X}_l$ is a finite set of actions available to layer $l$, each element $\xi_l \in \mathcal{X}_l$ of which contains the external and internal actions, i.e. $\xi_l = [a_l, b_l]$.
- $\Theta_{l,l+1}$ is the message set sent by layer $l$ to its upper layer $l+1$, where $\theta_{l,l+1} \in \Theta_{l,l+1}$ represents a message sent by layer $l$ to its upper layer $l+1$ (i.e. upward message).

- $\Theta_{l,l-1}$ is the message set sent by layer $l$ to its lower layer $l-1$, and $\theta_{l,l-1} \in \Theta_{l,l-1}$ represents a message sent by layer $l$ to its lower layer $l-1$ (i.e. downward message).
- $p$ is the transition probability function. $p(\boldsymbol{s}' \mid \boldsymbol{s}, \boldsymbol{\xi})$ is the probability of moving from state $\boldsymbol{s} \in \mathcal{S}$ to the state $\boldsymbol{s}' \in \mathcal{S}$ when layer $l \in \mathcal{L}$ performs action $\xi_l$. We assume that the transition model is stationary and independent of the stage (i.e. time).
- $R : \mathcal{S} \times \prod_{l=1}^{L} \mathcal{X}_l \mapsto \Re$ is the system stage reward function which has the form of $R(\boldsymbol{s}, \boldsymbol{\xi})$, i.e. the reward is determined by the state and actions in each layer.
- $\gamma$ is the discounted factor.

The framework of the layered MDP with information exchange for the foresighted cross-layer optimization problem is illustrated in Figure 3 (b). From this figure, we observe that the layer optimizer is not required to know other layers' state space, action space and dynamics models.

*B. Quality of service and upward message*

At the state $\boldsymbol{s}^k$, by deploying the internal actions, the WSTA can determine for each layer (i) the probability of the packet being successfully received at the destination; (ii) the amount of time it takes to transmit on average; and (iii) the cost associated with its transmission. The transmission result of whether a packet is successfully received, is represented by the average packet loss ratio (PLR) at layer $l$ at stage $k$, which is denoted by $\varepsilon_l^k\left(\boldsymbol{s}_{1\to l}^k, \boldsymbol{b}_{1\to l}^k\right)$ where $\boldsymbol{s}_{l'\to l''}^k = \left[s_{l'}^k, ..., s_{l''}^k\right]$ and $\boldsymbol{b}_{l'\to l''}^k = \left[b_{l'}^k, ..., b_{l''}^k\right]$ with $l' \le l''$. The average amount of time spent on transmitting one packet at layer $l$ at stage $k$ is denoted by $t_l^k\left(\boldsymbol{s}_{1\to l}^k, \boldsymbol{b}_{1\to l}^k\right)$. The aggregated transmission cost incurred by performing internal actions at layer $l$ is defined by $f_l^k\left(\boldsymbol{s}_{1\to l}^k, \boldsymbol{b}_{1\to l}^k\right) = \sum_{l'=1}^{l} \lambda_{l'}^b d_{l'}\left(s_{l'}^k, b_{l'}^k\right)$.

To compute the internal reward function $R_{in}\left(\boldsymbol{s}^k, \boldsymbol{b}^k\right)$, layer $L$ has to know the packet loss probability, the average amount of time for packet transmission and the transmission cost provided from the lower layers in stage $k$. We can define a message which captures this information from lower layers. This message is the QoS at layer $l$ which is defined as a three-tuple $Z_l^k = \left[\varepsilon_l^k, t_l^k, f_l^k\right]^T$. The QoS at layer $l$ represents the service layer $l$ provides to its upper layer $l+1$. Using the QoS, layer $l+1$ does not need to know the actions and dynamics at lower layers.

By knowing the QoS $Z_{L-1}^k$ provided from layer $L-1$, layer $L$ can be computed as $R_{in}\left(s_L^k, b_L^k \mid Z_{L-1}^k\right)$. In other words, the internal reward $R_{in}$ is independent of the states and actions in the lower layers, given the QoS $Z_{L-1}^k$ provided from layer $L-1$. More generally, the internal reward function $u\left(\boldsymbol{s}^k, \boldsymbol{b}^k\right)$ that we consider here has the following Markovian property:

**Definition 3. (Markovian internal reward function)**: The internal reward $R_{in}\left(\boldsymbol{s}, \boldsymbol{b}^k\right)$ at layer $L$ is Markovian if, given the QoS $Z_l^k$ at layer $l$, we have $R_{in}\left(\boldsymbol{s}^k, \boldsymbol{b}^k \mid Z_l^k\right) = R_{in}\left(\boldsymbol{s}_{l+1\to L}^k, \boldsymbol{b}_{l+1\to L}^k \mid Z_l^k\right)$. In other words, the QoS $Z_l^k$ is the sufficient statistics of the state $\boldsymbol{s}_{1\to l}^k$ and action $\boldsymbol{b}_{1\to l}^k$ in layers $\{1,...,l\}$ to compute $u\left(\boldsymbol{s}^k, \boldsymbol{b}^k\right)$.

Using the Markovian property of the internal reward function, layer $l$ only needs to report the possible QoS levels it can support to layer $l+1$, instead of reporting the action and dynamics. To further reduce the amount of messages to be reported to the upper layers, we define two possible relationships between the QoS levels: dominant and Pareto-equivalent. These relationships between the QoS levels enable the lower layers to provide the necessary QoS levels and will be essential to reduce the size of messages representing the QoS levels for all possible actions performed in layers $\{1,...,l\}$ for the upper layers given the states $\left[s_1^k,...,s_l^k\right]$.

**Definition 4. (Dominant)**: A QoS $Z_l^k$ is dominant with respect to another QoS $\tilde{Z}_l^k$, if $Z_l^k - \tilde{Z}_l^k \leq \mathbf{0}$ [3] and $Z_l^k \neq \tilde{Z}_l^k$.

The "dominant" relationship is denoted by $Z_l^k \overset{d.}{\leq} \tilde{Z}_l^k$. If the QoS $Z_l^k$ is obtained from a particular action profile $\boldsymbol{b}_{1\to l}^k$ and $\tilde{Z}_l^k$ is obtained by $\tilde{\boldsymbol{b}}_{1\to l}^k$ under the same state $\boldsymbol{s}_{1\to l}^k$, then we can also say that the action profile $\boldsymbol{b}_{1\to l}^k$ is dominant to the action profile $\tilde{\boldsymbol{b}}_{1\to l}^k$. Hence, we only need to consider the dominant QoS $Z_l^k$ and corresponding action $\boldsymbol{b}_{1\to l}^k$.

Similarly, we define the Pareto-equivalent relationship of two QoSs and their corresponding actions:

**Definition 5. (Pareto-equivalent)**: A QoS $Z_l^k$ is Pareto equivalent to another QoS $\tilde{Z}_l^k$, which is denoted by $Z_l^k \overset{p.e.}{=} \tilde{Z}_l^k$, if the following two conditions do not hold: $Z_l^k \overset{d.}{\leq} \tilde{Z}_l^k$ or $\tilde{Z}_l^k \overset{d.}{\leq} Z_l^k$.

Accordingly, we also say that the action profile $\boldsymbol{b}_{1\to l}^k$ is Pareto-equivalent to the action profile $\tilde{\boldsymbol{b}}_{1\to l}^k$.

In our cross-layer design framework, we consider the states and actions that will preserve the "dominant" relationship of the QoS levels. That is, the states and actions in each layer have the following property:

**Property 1 (Preservation of QoS)** : If $Z_l^k \overset{d.}{\leq} \tilde{Z}_l^k$, then $Z_{l+1}^k\left(s_{l+1}^k, b_{l+1}^k \mid Z_l^k\right) \overset{d.}{\leq} \tilde{Z}_{l+1}^k\left(s_{l+1}^k, b_{l+1}^k \mid \tilde{Z}_l^k\right)$ for any state $s_{l+1}^k$ and any action $b_{l+1}^k$.

In this paper, we consider the internal reward function at layer $L$ satisfies the following preservation property. The internal reward function in example 5 will be examined in Appendix II.

[3] $X \geq \mathbf{0}$ means every component of $X$ is greater than or equals 0.

**Property 2 (Preservation of internal reward function)** : If $Z_L^k \overset{d.}{\leq} \tilde{Z}_L^k$, then $R_{in}\left(s_L^k, Z_L^k\right) \leq R_{in}\left(s_L^k, \tilde{Z}_L^k\right)$.

The preservation of QoS means that the dominated QoS $\tilde{Z}_l^k$ provided by layer $l$ cannot bring a dominant QoS by performing any internal action at the upper layer. Hence, the dominated QoS $\tilde{Z}_l^k$ should not be reported to the upper layer. Hence, the preservation of dominant relationship reduces the information exchanged from lower layers to upper layers.

**Remark 1.** The goodput defined in the MAC layer (when $L = 2$) as in [4] satisfies these properties.

**Remark 2.** The utility defined in the APP layer such as the video quality also satisfies these properties [8].

To compute the internal reward function, we do not need to check all the possible combination of the internal actions from different layers. Since $Z_l^k$ is a sufficient statistics of the states $\boldsymbol{s}_{1\to l}^k$ and $\boldsymbol{b}_{1\to l}^k$, we only need to find the optimal frontier of the possible QoS set in each layer. We define the optimal frontier as follows.

**Definition 6. (Optimal frontier)**: The optimal frontier of the possible QoS set $\boldsymbol{\mathcal{Z}}_l^k$ at layer $l$ is a largest subset $\mathcal{Z}_l^k \subseteq \boldsymbol{\mathcal{Z}}_l^k$ with each element satisfying: for any $Z_l^k \in \mathcal{Z}_l^k$, there is no existing $\tilde{Z}_l^k \in \boldsymbol{\mathcal{Z}}_l^k$ such that $\tilde{Z}_l^k \overset{d.}{\leq} Z_l^k$.

Hence, each layer $l$ is only required to provide the QoS set $\mathcal{Z}_l^k$ that represents the optimal frontier instead of all the possible QoS levels (i.e. $\boldsymbol{\mathcal{Z}}_l^k$). Moreover, the optimal frontier is sufficient to maximize the internal reward function given the current state. This is summarized next.

**Proposition 1.** Maximizing the internal reward function $R_{in}\left(s_L^k, Z_L^k\right)$ over the optimal frontier $\mathcal{Z}_L^k$ is equivalent to maximizing over all the possible internal actions, i.e.

$$\max_{Z_L^k \in \mathcal{Z}_L^k} R_{in}\left(s_L^k, Z_L^k\right) = \max_{\boldsymbol{b} \in \prod_{l=1}^{L} \Delta_l^b} R_{in}\left(\boldsymbol{s}^k, \boldsymbol{b}^k\right). \tag{10}$$

From the description about the preservation of dominant relationship and internal reward function, the equivalence in Eq. (10) can be easily shown. We omit this proof to save the space.

By reporting the optimal frontier from the lower layer to upper layer, we reduce the size of the message (containing only $\mathcal{Z}_l^k$ instead of $\boldsymbol{\mathcal{Z}}_l^k$) passed from layer $l$ to $l+1$. For the layered MDP, the upward message from layer $l$ to layer $l+1$ becomes $\theta_{l,l+1}^k = \mathcal{Z}_l^k$. We should also note that the message $\theta_{l,l+1}^k$ encapsulates the states and actions performed at layer $l$ and below, while having a format that is independent of the protocols and algorithms implemented at those layers.

### C. Layered value iteration and downward messages

In Section IV.B, we illustrate the value iteration for the centralized cross-layer optimization. Using our layered MDP framework, we propose a layered value iteration algorithm by allowing the information exchange between adjacent layers.

As defined in Definition 2, each layer in the layered MDP is regarded as an autonomous entity that performs its own actions. However, the layers can cooperate via the information exchange to find the optimal state-value function $V^*(\boldsymbol{s})$. By decomposing the value iteration in Eq. (9), we can obtain the following theorem.

**Theorem 1:** The state-value function $V^*(\boldsymbol{s})$ corresponding to the optimal policy can be obtained using a layered value iteration algorithm. At iteration $n$, each layer performs a sub-value iteration which is given in Table 1.

Table 1. Value iteration at each layer.

| Layer | Value iteration form at iteration $n$ | |
|---|---|---|
| $L$ | $V_{n,L-1}^*\left(\boldsymbol{s}_{1\to L\text{-}1}'\right) = \max_{a_L \in \Delta_L^a, Z_L \in \mathcal{Z}_L} \left[R_{in}(s_L, Z_L) - \lambda_L c_L(s_L, a_L) + \gamma \sum_{s_L \in \mathcal{S}_L} p\left(s_L' \mid \boldsymbol{s}_{1\to L-1}', s_L, a_L, Z_L\right) V_{n-1,L}^*\left(\boldsymbol{s}_{1\to L}'\right)\right]$ | (11) |
| $l \in \{2,...,L-1\}$ | $V_{n,l-1}^*\left(\boldsymbol{s}_{1\to l\text{-}1}'\right) = \max_{a_l \in \Delta_l^a}\left[-\lambda_l c_l(s_l, a_l) + \sum_{s_l' \in \mathcal{S}_l} p\left(s_l' \mid \boldsymbol{s}_{1\to l-1}', s_l, a_l\right) V_{n,l}^*\left(\boldsymbol{s}_{1\to l}'\right)\right]$ | (12) |
| 1 | $V_{n,L}^*(\boldsymbol{s}_{1\to L}) = \max_{a_1 \in \Delta_1^a}\left[-\lambda_1 c_1(s_1, a_1) + \sum_{s_1' \in \mathcal{S}_1} p\left(s_1' \mid \boldsymbol{s}_1, a_1\right) V_{n,1}^*\left(\boldsymbol{s}_1'\right)\right]$ | (13) |

Note: $V_{n,l}^*(\boldsymbol{s}_{1\to l})$ is the state-value function of state $\boldsymbol{s}_{1\to l}$ used in the sub-value iteration at layer $l$ at iteration $n$. $V_{n,L}^*(\boldsymbol{s}_{1\to L}) = V_n^*(\boldsymbol{s})$.

The proof is given in Appendix I.

The layered value iteration is performed as follows: at each iteration $n$, layer $L$ performs the sub-value iteration as in Eq. (11) to obtain the state-value function $V_{n,L-1}^*\left(\boldsymbol{s}_{1\to L\text{-}1}'\right)$ which services as future state-value function at layer $L-1$. Then, in general, layer $l$ performs the sub-value iteration as in Eq. (12) based on the future state-value function from layer $l+1$ to generate $V_n^*\left(\boldsymbol{s}_{1\to l-1}'\right)$. Finally, layer 1 performs the sub-value iteration as in Eq. (13) to generate the state-value function $V_{n,L}^*(\boldsymbol{s}_{1\to L})$ which is $V_n^*(\boldsymbol{s})$ as in the centralized value iteration.

Then the message exchanged from layer $l+1$ to layer $l$ is $\theta_{l+1,l} = \left\{V_{n-1}^*\left(\boldsymbol{s}_{1\to l}'\right)\right\}$. The upward and downward message exchange is presented in Table 2.

Table 2. Message exchanges between layers at iteration $n$ .

| Layer | Upward Message $\theta_{l,l+1}$ | | Downward Message $\theta_{l,l-1}$ | |
|---|---|---|---|---|
| $L$ | $\varnothing$ | None | $\left\{V_n^*\left(\boldsymbol{s}_{1\to L-1}\right)\right\}$ | Expected future reward at layer $L-1$ |
| $l \in \{2,...,L-1\}$ | $\mathcal{Z}_l$ | QoS level set provided to layer $l+1$ | $\left\{V_n^*\left(\boldsymbol{s}_{1\to l-1}\right)\right\}$ | Expected future reward at layer $l-1$ |
| 1 | $\mathcal{Z}_1$ | QoS level set provided to layer 2 | $\varnothing$ | None |

### D. *Advantages of layered value iteration*

In this section, we highlight the advantages of the proposed layered value iteration compared to the centralized value iteration illustrated in Section IV.B.

We first compare the complexity of these two algorithms for solving the foresighted cross-layer optimization problem. Due to the mixed actions performed at each layer, in the centralized value iteration, the central optimizer needs to check the whole space $\prod_{l=1}^{L} \Delta_l^a \times \Delta_l^b$ to find the optimal mixed actions for each state $\boldsymbol{s}$ at each iteration. However, in the proposed layered value iteration, layer $L$ needs to check the action space $\Delta_L^a$ for $|\mathcal{Z}_l||\mathcal{S}_{1\to L-1}|$ times, layer $l \in \{2,...,L-1\}$ needs to check $\Delta_l^a$ for $|\mathcal{S}_{1\to l-1}|$ times and layer 1 needs to check $\Delta_1^a$ once for each state $\boldsymbol{s}$. Since the state space is finite, the action space the layered MDP has to check is smaller than that in the centralized MDP. Hence, we have the following remark.

**Remark 3.** The proposed layered value iteration shown in Table 1 has lower complexity than the traditional centralized value iteration shown in Eq. (9).

As discussed in Section IV, the central optimizer is required to completely know the dynamics model (i.e. states, transition probability) and possible internal and external actions of all the layers which are protocol-dependent. Hence, the mechanism of information exchange between the central optimizer and the layers is also protocol-dependent. In the proposed algorithm, however, the centralized value iteration shown in Eq. (9) is decomposed into multiple sub-value iteration procedures each of which is accordingly solved by one layer. From the sub-value iteration for each layer shown in Table 1 and the message exchange between layers shown in Table 2, we note that our proposed layered MDP framework has the following advantages:

First, to perform the sub-value iteration given the information exchanged between layers, each layer is only required to know its own internal and external actions and transition probabilities (corresponding to the dynamics models) but it is not required to know the actions and transition probabilities of other layers, given the information exchanged between layers.

Second, the format (i.e. QoS frontier for upward message and state-value function for downward message) of the messages exchanged between layers is independent of the protocols deployed in each layers although the content (i.e. QoS frontier is resulted from the performed internal actions and state-value function is resulted from the external actions) of the messages characterizes the dynamics and performed actions at each layer.

Third, since the format of messages is independent of the protocols, the proposed layered MDP framework will not be changed even if the protocols at a particular layer are upgraded.

We compare our proposed layered cross-layer optimization to the existing layered architecture without cross-layer optimization and traditional cross-layer optimization to highlight the merits of our proposed layered MDP framework for the cross-layer optimization. The comparison results are shown in Figure 4.

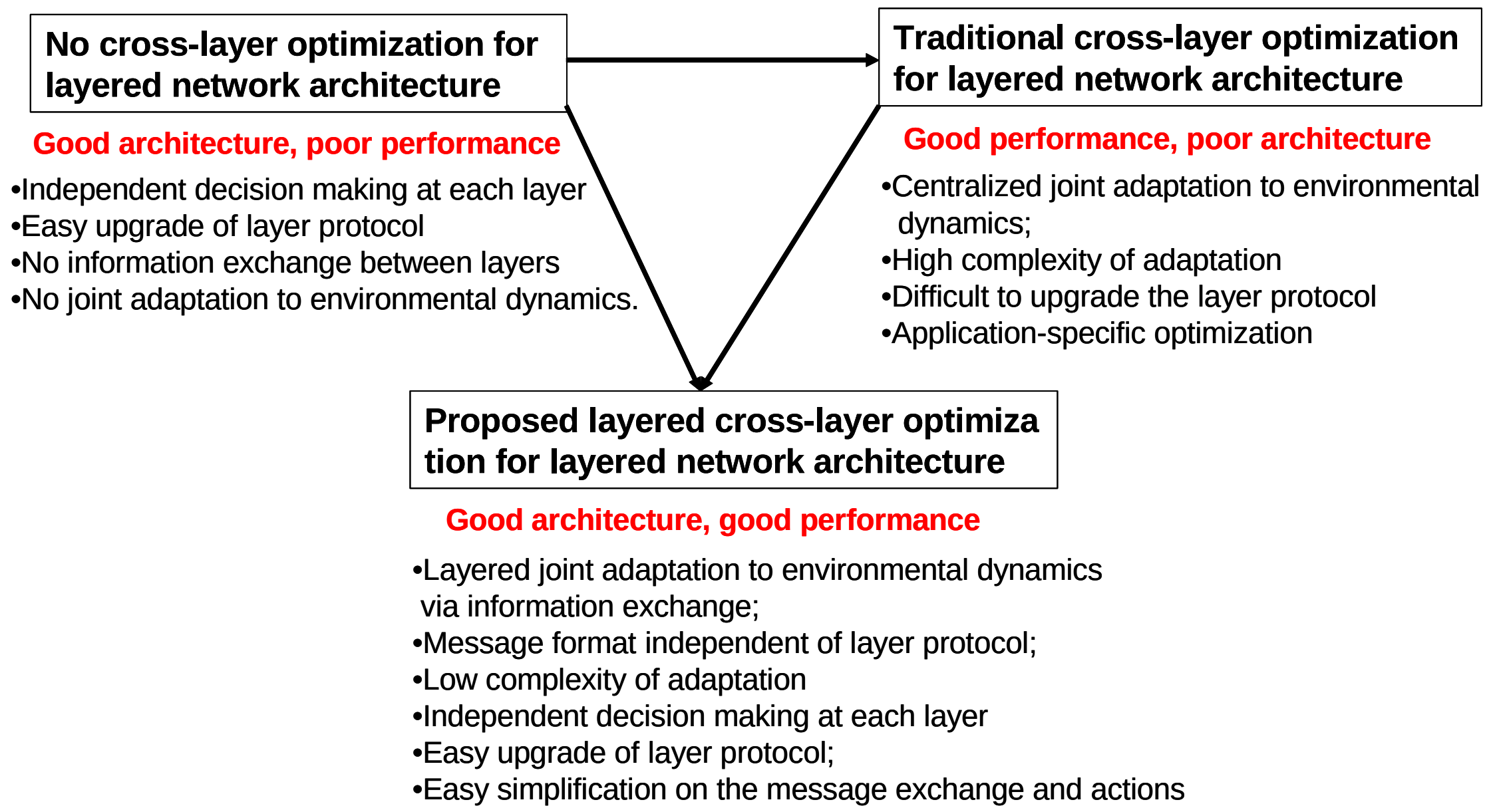

Figure 4. Comparison of layered networked architecture without cross-layer optimization, traditional cross-layer optimization and proposed layered cross-layer optimization

### *E. Simplifications of the layered MDP framework*

The proposed layered MDP framework explicitly considers the selection of the internal and external policies at each layer. Actually, many existing cross-layer approaches are ad-hoc simplified versions of the proposed framework. For example, many quality-oriented cross-layer optimization approaches discussed in Section II.A simplify the layered MDP framework by limiting the information exchange between layers, while many QoS-oriented approaches ignores the internal and external actions in the APP layers.

In this section, we illustrate how the proposed framework can be simplified by limiting the information exchange between layers or restricting the internal and external actions, and determine the resulting performance penalty obtained by these simplified cross-layer optimization solutions.

**Simplification 1**: Assume that the lower layers $\{1,...,L-1\}$ use constant external and internal actions. There are no upward and downward messages exchanged between layer $L$ and the lower layers. In this case, layer $L$ models the QoS $Z_{L-1}$ provided by the lower layers as a random variable due to the unknown information about the lower layers.

Based on the value iteration form shown in Table 1, layer $L$ performs the value iteration as follows:

$$V_n^*(s_L) = \max_{\xi_L \in \Delta_l^a \times \Delta_l^b} E_{Z_{L-1}} \left[ u(s_L, b_L, Z_{L\text{-}1}) - \lambda_L c_L(s_L, a_L) + \gamma \sum_{s_L \in \mathcal{S}_L} p\left(s_L' \mid s_L, a_L, b_L, Z_{L\text{-}1}\right) V_{n-1}^*\left(s_L'\right) \right] \quad (14)$$

The difference between Eqs. (14) and (11) is the following: first, the maximization is performed on the expected value of the current stage reward and future reward with respect to the QoS $Z_{L-1}$; second, the obtained optimal value is $V_n^*(s_L)$ instead of $V_{n,L-1}^*\left(\boldsymbol{s}_{1\to L\text{-}1}'\right)$.

The solution proposed in [8] represents this type of simplification of our framework. However, we should note that this simplification is achieved at the expense of performance loss, since the cross-layer adaptation in the lower layers is neglected.

**Proposition 2.** Simplification 1 results in suboptimal performance.

Proof: Assume that $\pi^*$ is the optimal policy obtained by the value iteration algorithm shown in Table 1 for the original foresighted cross-layer optimization problem (i.e. the one without simplification). On the other hand, the value iteration for the simplified version of the cross-layer problem has the optimal policy $\pi_L^{simplified}$. Then we can construct a policy $\pi^{simplified} = \left[\xi_1, ..., \xi_{L-1}, \pi_L^{simplified}\right]$ with $\xi_l$ being the action performed at layer $l$ no matter what state it is in. Hence $\pi^{simplified}$ is a policy for the original cross-layer optimization. Since $\pi^*$ is an optimal policy, we have $V\left(\pi^*, \boldsymbol{s}\right) \geq V\left(\pi^{simplified}, \boldsymbol{s}\right), \forall \boldsymbol{s}$. □

**Simplification 2**: Assume that there are no external and internal actions performed at layer $L$.

Layer $L$ generates the message $V_{n-1}^*\left(\boldsymbol{s}_{1\to L\text{-}1}'\right)$ as follows:

$$V_{n-1}^*\left(\boldsymbol{s}_{1\to L\text{-}1}'\right) = \max_{Z_{L-1} \in \mathcal{Z}_{L\text{-}1}} \left[ u(s_L, Z_{L\text{-}1}) + \gamma \sum_{s_L \in \mathcal{S}_L} p\left(\boldsymbol{s}_L' \mid \boldsymbol{s}_L, Z_{L\text{-}1}\right) V_{n-1}^*\left(\boldsymbol{s}'\right) \right] \quad (15)$$

The difference between Eqs. (15) and (11) is that the maximization in Eq. (15) is only with respect to the QoS levels provided by the lower layers. Since the QoS level set $\mathcal{Z}_{L\text{-}1}$ is computed by the layer $L-1$, the optimization in Eq. (15) is often performed in layer $L-1$ by allowing layer $L-1$ to access the state and transition probability at layer $L$. If the traffic model at layer $L$ is known, the transition probability $p\left(\boldsymbol{s}_L' \mid \boldsymbol{s}_L, Z_{L\text{-}1}\right)$ can be easily computed [4], e.g. Poisson packet arrival. The research in [4][5][6] performs this type of simplification. This simplification does not take into account the external and internal actions at layer $L$, and often leads to suboptimal performance.

**Proposition 3.** Simplification 2 results in the suboptimal performance.

The proof is similar to the one for Proposition 2 and hence, it is omitted here.

The optimal policy for the simplified version of the layered MDP is $\pi^{simplified}$ and the optimal policy for the original layered MDP (without simplification) is $\pi^*$. We can deploy both policies on-line and compute stage reward over time. Then the performance loss due to the simplification can be computed as

$$\Delta R = \frac{1}{K}\sum_{k=1}^{K}\left(R\left(\tilde{\boldsymbol{s}}^k, \pi^{simplified}\left(\tilde{\boldsymbol{s}}^k\right) \mid \boldsymbol{s}^0\right) - R\left(\boldsymbol{s}^k, \pi^*\left(\boldsymbol{s}^k\right) \mid \boldsymbol{s}^0\right)\right) \tag{16}$$

where $\boldsymbol{s}^0$ is the initial state for both policies and $K$ is the number of stages under consideration.

In Section VII.G, we further illustrate how much performance loss is incurred by these two types of simplifications.

# VI. ON-LINE ADAPTATION

In Section V, we proposed a layered value iteration solution for the foresighted cross-layer optimization, which can find the optimal internal and external policies in an *off-line* fashion when the models for the environment dynamics are known. In this section, we extend the layered MDP framework to perform on-line adaptation for the case in which the models are unknown.

## A. *Message exchange for on-line adaptation*

From the upward message exchange shown in Table 2 between layers, each layer $l \in \{1,...,L-1\}$ provides the optimal frontier $\mathcal{Z}_l$ to its upper layer. Layer $L$ (e.g. APP layer) chooses the optimal QoS $Z_{L\text{-}1}^* \in \mathcal{Z}_{L-1}$ by performing the value iteration. From the optimal QoS $Z_{L\text{-}1}^*$, each layer can automatically obtain the optimal internal actions. From this perspective, the optimal internal actions are selected in a similar manner to the application-centric on-line adaptation [2].

From the downward message exchange shown in Table 2 between layers, each layer $l \in \{2,...,L\}$ provides the message $\{V_l(\boldsymbol{s}_{1\to l-1})\}$ to layer $l-1$. The message $\{V_l(\boldsymbol{s}_{1\to l})\}$ serves as the expected future reward for the value iteration at layer $l$. Based on $\{V_l(\boldsymbol{s}_{1\to l})\}$, layer $l$ can select its own external action that maximizes not only its myopic reward, but also the future reward.

### B. On-line adaptation using actor-critic learning

The layered value iteration proposed in Section V assumes that the dynamics models, i.e. transition probability, at each layer are known a priori. When the models are unknown, these models can be learned using learning techniques [28]. To highlight the advantage of the proposed layered MDP framework, we use for illustration the actor-critic reinforcement learning algorithm [34][35] for the on-line transmission strategy adaptation. The actor-critic algorithm separately updates the policy and the state-value function for each state. The *policy structure* is used to select actions at each state and is called the *actor*. The *state-value function* is used to criticize the actions selected by the actor and is called the *critic*. We briefly discuss the state-value function and policy update used in the actor-critic algorithm. The algorithm's details about the algorithm can be found in [35].

#### 1) State-value function update

During the on-line adaptation, the state-value function $V(\boldsymbol{s})$ is unknown and must be estimated on-line. Recall the recursive form in Eq. (8) for the computation of the value function of policy $\pi$. When performing action $\boldsymbol{\Psi}^k$ [4], we can update the state-value function, given the current reward $R(\boldsymbol{s}^k, \boldsymbol{\Psi}^k)$ as follows:

$$V^{k+1}(\boldsymbol{s}^k) \leftarrow V^k(\boldsymbol{s}^k) + \alpha\left[R(\boldsymbol{s}^k, \boldsymbol{\Psi}^k) + \gamma V^k(\boldsymbol{s}^{k+1}) - V^k(\boldsymbol{s}^k)\right] \quad (17)$$

where $\alpha$ is a positive step-size parameter and $V^k(\cdot)$ and $V^{k+1}(\cdot)$ are the estimated future rewards for stage $k$ and stage $k+1$, respectively. Since the real future reward is unknown, $V^k(\boldsymbol{s}^{k+1})$ is used instead to update the state-value function. The update procedure for the value function in the actor-critic learning algorithm is performed in the *state-value function* update module illustrated in Figure 5.

#### 2) Policy update

The state-value function $V^k(\cdot)$ is used to criticize the selected action. After each action selected by the actor, the critic evaluates the selected action at current state $\boldsymbol{s}^k$ to determine whether the value function at the current state performs better or worse than expected. This evaluation can be defined as the time-difference error as follows:

$$\delta^k = R(\boldsymbol{s}^k, \boldsymbol{\Psi}^k) + \gamma V^k(\boldsymbol{s}^{k+1}) - V^k(\boldsymbol{s}^k) \quad (18)$$

If the error $\delta^k$ is positive, it means that the tendency to select action $\boldsymbol{\Psi}^k$ should be strengthened in the future, while if it is negative, the tendency to select $\boldsymbol{\Psi}^k$ should be weakened.

[4] We consider here the real actions instead of the mixed actions. However, the action generated by the actor is mixed.

To generate the action, the actor defines a value $\rho(\boldsymbol{s},\boldsymbol{\Psi})$ at state $\boldsymbol{s}$ for each action $\boldsymbol{\Psi}$ to indicate the tendency to select that action. Then the actor generates the action according to the Gibbs softmax method [35]:

$$\boldsymbol{\xi}(\boldsymbol{s},\boldsymbol{\Psi}) = \frac{e^{\rho(\boldsymbol{s},\boldsymbol{\Psi})}}{\sum_{\boldsymbol{\Psi}' \in \prod_{l=1}^{L} \mathcal{X}_l} e^{\rho(\boldsymbol{s},\boldsymbol{\Psi}')}} \tag{19}$$

where $\boldsymbol{\xi}(\boldsymbol{s},\boldsymbol{\Psi})$ represents the probability of performing action $\boldsymbol{\Psi}$ at state $\boldsymbol{s}$.

The strengthening and weakening of the action can then be implemented by increasing or decreasing the tendency as follows:

$$\rho\left(\boldsymbol{s}^k,\boldsymbol{\Psi}^k\right) \leftarrow \rho\left(\boldsymbol{s}^k,\boldsymbol{\Psi}^k\right) + \beta\delta^k\left(1 - \boldsymbol{\xi}\left(\boldsymbol{s}^k,\boldsymbol{\Psi}^k\right)\right) \tag{20}$$

where $\beta$ is a positive step-size parameter and reflects the learning rate for the tendency update. The policy update is performed in the *policy* module in Figure 5.

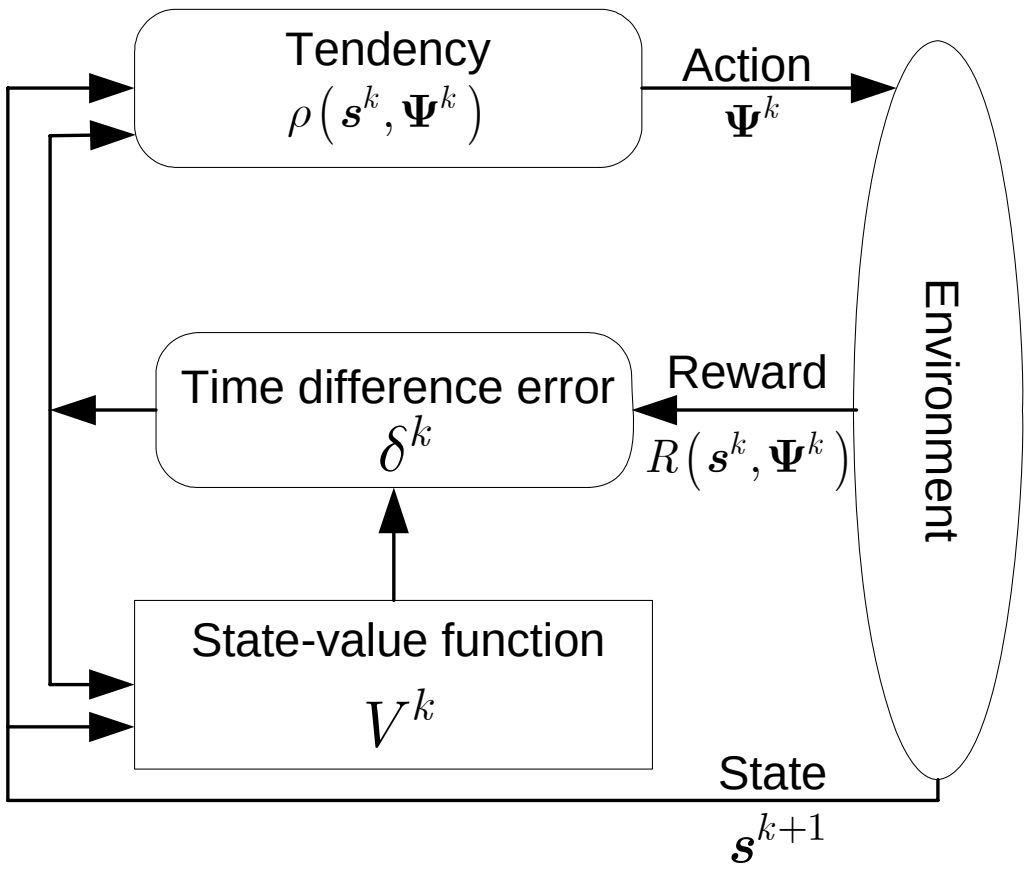

Figure 5.Actor-critic learning structure based on [35]

### C. *On-line adaptation using layered learning*

Based on the actor-critic learning algorithm, we develop a layered actor-critic learning algorithm which takes into account the current layered network architecture. In this layered learning algorithm, each layer has its own actor and critic to select the action and criticize the selected action.

*1) State-value update*

Recall the value iteration at layer $l \in \{1,...,L\}$. We can define the time-difference error at layer $l$ as

$$\delta_l^k = \begin{cases} u\left(s_L^k, A_L^k, Z_L^k\right) + \gamma V^k\left(\boldsymbol{s}^{k+1}\right) - V_{L-1}^k\left(\boldsymbol{s}_{1\to L-1}^{k+1}\right) & l = L \\ -\lambda_l^c c_l\left(A_l^k\right) + V_l^k\left(\boldsymbol{s}_{1\to l}^{k+1}\right) - V_{l-1}^k\left(\boldsymbol{s}_{1\to l-1}^{k+1}\right) & l = 2,...,L-1 \\ -\lambda_1^c c_1\left(A_1^k\right) + V_1^k\left(s_1^{k+1}\right) - V^k\left(\boldsymbol{s}^k\right) & l = 1 \end{cases}. \tag{21}$$

Then, $V_l^{k+1}\left(\boldsymbol{s}_{1\to l}^{k+1}\right)$ is updated as

$$V_l^{k+1}\left(\boldsymbol{s}_{1\to l}^{k+1}\right) \leftarrow V_l^{k+1}\left(\boldsymbol{s}_{1\to l}^{k+1}\right) + \alpha\delta_{l+1}^k, l = 1,...,L-1. \tag{22}$$

The state-value function $V^{k+1}\left(\boldsymbol{s}^k\right)$ is updated as

$$V^{k+1}\left(\boldsymbol{s}^k\right) \leftarrow V^k\left(\boldsymbol{s}^k\right) + \alpha\left[R_{in}\left(s_L^k, Z_L^k\right) - \sum_{l=1}^{L}\lambda_l^c c_l\left(s_l^k, A_l^k\right) + \gamma V^k\left(\boldsymbol{s}^{k+1}\right) - V^k\left(\boldsymbol{s}^k\right)\right], l = 1,...,L-1 \tag{23}$$

*2) Policy update*

Given the state at each layer, the internal actions are independent of the environmental dynamics. In this learning algorithm, the state $\boldsymbol{s}$ is assumed to be known by each layer. From Section V, we know that the optimal frontier $\mathcal{Z}_L$ only depends on the state $\boldsymbol{s}$, and hence, layer $L$ can select the optimal QoS $Z_L \in \mathcal{Z}_L$. The tendency at layer $L$ is updated using the time-difference error $\delta_L^k$ to strengthen or weaken the currently selected action, as follows:

$$\rho\left(\boldsymbol{s}^k, A_L^k, Z_L^k\right) \leftarrow \rho\left(\boldsymbol{s}^k, A_L^k, Z_L^k\right) + \beta\delta_l^k\left(1 - \xi\left(\boldsymbol{s}^k, A_L^k, Z_L^k\right)\right) \tag{24}$$

Similarly, the tendency at layer $l$ is updated as

$$\rho\left(\boldsymbol{s}^k, A_l^k\right) \leftarrow \rho\left(\boldsymbol{s}^k, A_l^k\right) + \beta\delta_l^k\left(1 - a_l\left(\boldsymbol{s}^k, A_l^k\right)\right), l = 1,...,L-1 \tag{25}$$

The action is generated similarly to Eq. (19). The layered learning algorithm is portrayed in Figure 6.

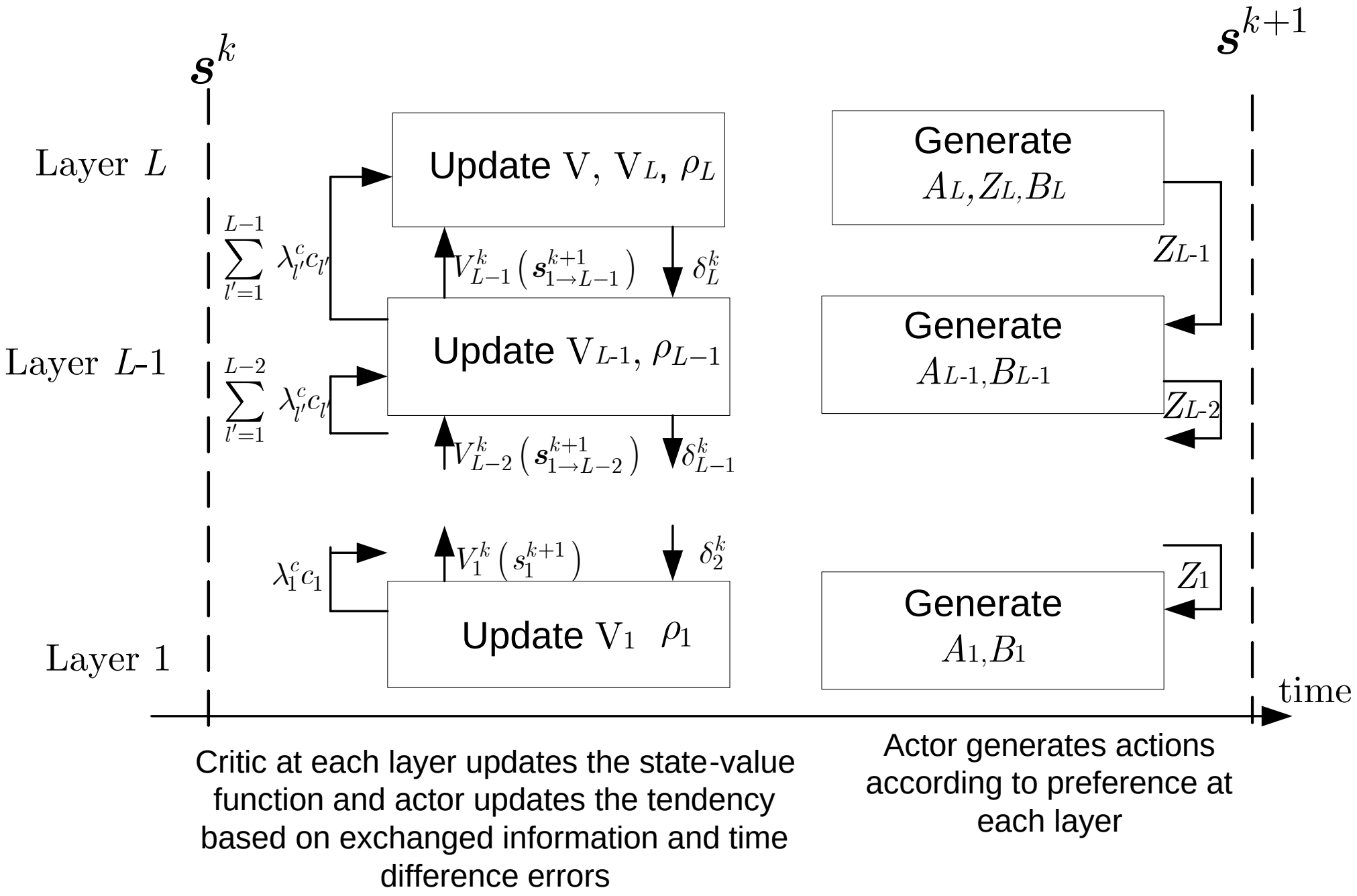

Figure 6.Proposed layered actor-critic learning procedure

# VII. Simulation Results for The Illustrative Example

In this section, we use example 1 to illustrate the proposed cross-layer design framework. We first discuss the states, actions and dynamics model used at each layer. Then we provide simulation results to illustrate the merits of our proposed layered MDP framework for cross-layer optimization.

## A. Application layer models

In the APP layer, we assume that the WSTA deploys a delay sensitive application. The date of the application layer is packetized with an average packet length $\ell$. In this paper, we consider an application where the application packets have a hard delay deadline, i.e. the packets will expire after $J$ stages after they are ready for transmission. Then, we can define the state of the APP layer as $s_3^k = \left[s_{3,1}^k, \cdots, s_{3,J}^k\right]^T$, where $s_{3,j}^k\ (1 \leq j \leq J)$ is the number of packets waiting for transmission which have a remaining life time of $j$ stages.

We denote by $Y_3^k$ the random number of packets with life time $J$ arriving into the buffer at the beginning of stage $k$. The average value of $Y_3^k$ is determined by the external action (e.g. the source coding parameter adaptation) in the APP layer. For simplicity, we refer to the external action in the APP layer as the average number of incoming packets, i.e. $E\left[Y_3^k\right] = A_3^k$. The probability mass function (PMF) of the random variable $Y_3^k\left(A_3^k\right)$ is assumed to be independent at each stage and denoted by $\left\{P\left(Y_3^k = y \mid A_3^k\right); y \in \mathbb{N}\right\}$.

Given the QoS $Z_3^k$, the number of packets transmitted is computed as

$$v_3^k\left(Z_3^k\right) = \left\lfloor \frac{\eta}{t_3^k}\left(1 - \varepsilon_3^k\right) \right\rfloor \tag{26}$$

where $\eta$ is the length of one stage in seconds. The state transition is characterized by

$$\begin{bmatrix} s_{3,1}^{k+1} \\ \vdots \\ s_{3,j}^{k+1} \\ \vdots \\ s_{3,J}^{k+1} \end{bmatrix} = \begin{bmatrix} s_{3,2}^k - \max\left(v_3^k\left(Z_3^k\right) - s_{3,1}^k, 0\right) \\ \vdots \\ s_{3,2}^k - \max\left(v_3^k\left(Z_3^k\right) - \sum_{m=1}^{j-1} s_{3,m}^k, 0\right) \\ \vdots \\ Y_3^k\left(A_3^k\right) \end{bmatrix} \tag{27}$$

The gain for the delay-sensitive application is defined here as

$$g\left(s_3^k, Z_3^k\right) = v_3^k\left(Z_3^k\right) - \lambda_g \min\left\{s_{3,1}^k - v_3^k\left(Z_3^k\right), 0\right\}, \tag{28}$$

where $\lambda_g$ is the parameter to trade-off the received packets and lost packets.

The state transition probability is computed as

$$p\left(s_{APP}^{k+1} \mid s_3^k, A_3^k, Z_L^k\right) = \begin{cases} P\left(M_3^k = m \mid A_3^k\right) & \textit{if } s_3^{k+1} \textit{ satisfies eq.} (27) \textit{ and } M_3^k = m \\ 0 & o.w. \end{cases} . \tag{29}$$

### B. MAC layer model

In MAC layer, each WSTA requests spectrum access by performing the external actions $A_2^k$ which can be the resource requests values (e.g. taxation). The MAC layer state $s_2^k$ is the number of time slots allocated in the current stage ( $N$ time slots in total per stage). By taking external action $A_2^k$, the transition probability is $p\left(s_2^{k+1} \mid s_2^k, A_2^k\right)$ and the cost introduced is $c_2\left(s_2^k, A_2^k\right) = A_2^k$.

In the MAC layer, the WSTA can perform ARQ to enhance the QoS provided to the application layer. Hence, the internal action can be $B_2^k \in \{0, ..., N_{\max}\}$ where $N_{\max}$ is the maximum retry limit and $B_2^k$ is the retry limit. Given the QoS provided from the PHY layer, say $Z_1^k = \left(\varepsilon_1^k, t_1^k\right)$, if the internal action $B_2^k$ is performed, then the QoS obtained in the MAC layer becomes

$$Z_2^k = \left(\varepsilon_2^k, t_2^k\right) = \left(\left(\varepsilon_1^k\right)^{B_2^k+1}, \frac{\left(1-\left(\varepsilon_1^k\right)^{B_2^k}\right)t_1^k}{\left(1-\varepsilon_1^k\right)s_2^k}\right) \tag{30}$$

It is easy to show that, if $Z_1^k \overset{d.}{<} \tilde{Z}_1^k$, then $Z_2^k \overset{d.}{<} \tilde{Z}_2^k$ for any internal action $B_2^k$, which means that the preservation of QoS property defined in Section IV is satisfied.

### C. Physical layer model

Similar to the model used in [25][26], we assume that the wireless channel gain experienced by a WSTA can be modeled as a discrete time FSMC. The state $s_1^k$ in the PHY layer is the channel gain. The WSTA is able to adapt its modulation scheme and power allocation for different states. We define the adaptation of the modulation $m \in \mathfrak{M}$ where $\mathfrak{M}$ is the set of possible modulation scheme, and power allocation $\sigma \in \mathcal{P}$ where $\mathcal{P}$ is the set of possible power allocation, as the internal actions, $B_1^k = (m, \sigma)$. As shown in [6], the PHY layer state can be determined by partitioning the possible received channel gain into $r+1$ disjoint regions $\mathbb{R}_0, ..., \mathbb{R}_r$ by boundary points $\Gamma_0, ..., \Gamma_{r+1}$, where $\mathbb{R}_i = [\Gamma_i, \Gamma_{i+1})$ and $\Gamma_0 < \Gamma_1 < ... < \Gamma_{r+1}$. The PHY layer is said to be in the state $s_1^k = \tilde{\Gamma}_i$ where $\tilde{\Gamma}_i$ is the representative channel gain if the real channel gain is in the region $\mathbb{R}_{i-1}$. Similar to [26], the channel gain is assumed to be a Rayleigh fading channel, denoted by $\Upsilon$, which is exponentially distributed with the following probability density function:

$$p_{\Upsilon}(\mu) = \frac{1}{\bar{\mu}} \exp\left(-\frac{\mu}{\bar{\mu}}\right), \mu \geq 0 \tag{31}$$

where $\bar{\mu}$ is the average channel gain. The state transition at the PHY layer is computed as

$$p\left(s_1^{k+1} \mid s_1^k\right) = \begin{cases} \mathcal{N}\left(\tilde{\Gamma}_{i+1}\right)\dfrac{T_p}{\omega_i} & s_1^k = \tilde{\Gamma}_i, s_1^{k+1} = \tilde{\Gamma}_{i+1} \\ \mathcal{N}\left(\tilde{\Gamma}_i\right)\dfrac{T_p}{\omega_i} & s_1^k = \tilde{\Gamma}_i, s_1^{k+1} = \tilde{\Gamma}_{i-1} \\ 0 & o.w. \end{cases} \tag{32}$$

Where $\mathcal{N}(\mu) = \sqrt{2\pi\mu / \bar{\mu}} f_d \exp(-\mu / \bar{\mu})$, $\omega_i = \exp(-\Gamma_i / \bar{\mu}) - \exp(-\Gamma_{i+1} / \bar{\mu})$, $T_p$ is the transmission time for one packet and $f_d$ is the maximum Doppler frequency.

### D. *Stage reward function*

In this example, the internal reward function is given by $R_{in}\left(s_3^k, Z_3^k\right) = v_3^k\left(Z_3^k\right) - \lambda_g \min\left\{s_{3,1}^k - v_3^k\left(Z_3^k\right), 0\right\} - Z_3^k(3)$ where $Z_3^k(3)$ is the third element in $Z_3^k$ which is the cost of power allocation. In Appendix II, we prove that the internal reward function $R_{in}\left(s_3^k, Z_3^k\right)$ is a non-decreasing function of $Z_3^k$, i.e. $R_{in}\left(s_3^k, Z_3^k\right) \geq R_{in}\left(s_3^k, \tilde{Z}_3^k\right)$ if $Z_3^k \overset{d.}{\leq} \tilde{Z}_3^k$.

### E. *Verification of the optimality of layered value iteration*

Table 3. Parameters used for the simulation at the various layers

| Layer | Parameter | Value |
|---|---|---|
| PHY layer | Channel gain model parameters | $f_d = 50\text{Hz}$, $T_p = 0.8\text{ms}$, $s_1 \in [-8, -6, ..., 8]\text{dB}$ |
| | Modulation level | $m = 1, ..., 4$ (BPSK, QPSK, 8PSK, 16PSK) |
| | Power allocation | $\sigma \in [0, 0.2, ..., 2]\,\text{Watt}$ |
| | Packet loss probability | $\varepsilon_1 = 1 - (1 - BER)^{\eta}$<br>$BER(s_1, m, \sigma) = erfc\left(\kappa\sigma\Gamma_{s_1} \sin\left(\frac{\pi}{2^m}\right)\right), \kappa = 283.5$ |
| | Transmission time per packet | $T_p / m$ |
| | Internal Lagrange multiplier | $\lambda_1^b = 1$ |
| MAC layer | Maximum life time | $J = 2$ |
| | MAC state | $s_2 \in \{0.1, 0.5, 1\}$ |
| | Maximum retransmission limit | $N_{\max} = 5$ |
| | External Lagrange multiplier | $\lambda_2^a = 1$ |
| | Competition bids (external action) | $A_2 \in \{0, 1\}$ |
| APP layer | APP state | $s_3 \in \{(0,0), ..., (4,4)\}$ |
| | External action | $A_3 \in \{1, 2, 3\}$ |
| | Lagrange multiplier | $\lambda_g = 0.1$ |

In this simulation, we compare the performance of the centralized value iteration and our proposed layered value iteration. Through this comparison, we will verify that the proposed layered value iteration also optimally solves the cross-layer optimization problem defined in Section IV. The parameters for the APP, MAC and PHY layers are shown in Table 3. The state-value functions $V^*(\mathbf{s})$ resulting from the centralized value iteration and proposed layered value iteration are shown in Figure 7. From this figure, we observe that the state-value functions computed based on both the algorithms are the same, which means that our proposed layered value iteration algorithm achieves the same performance as the centralized one, i.e. optimally finding the cross-layer transmission strategies.

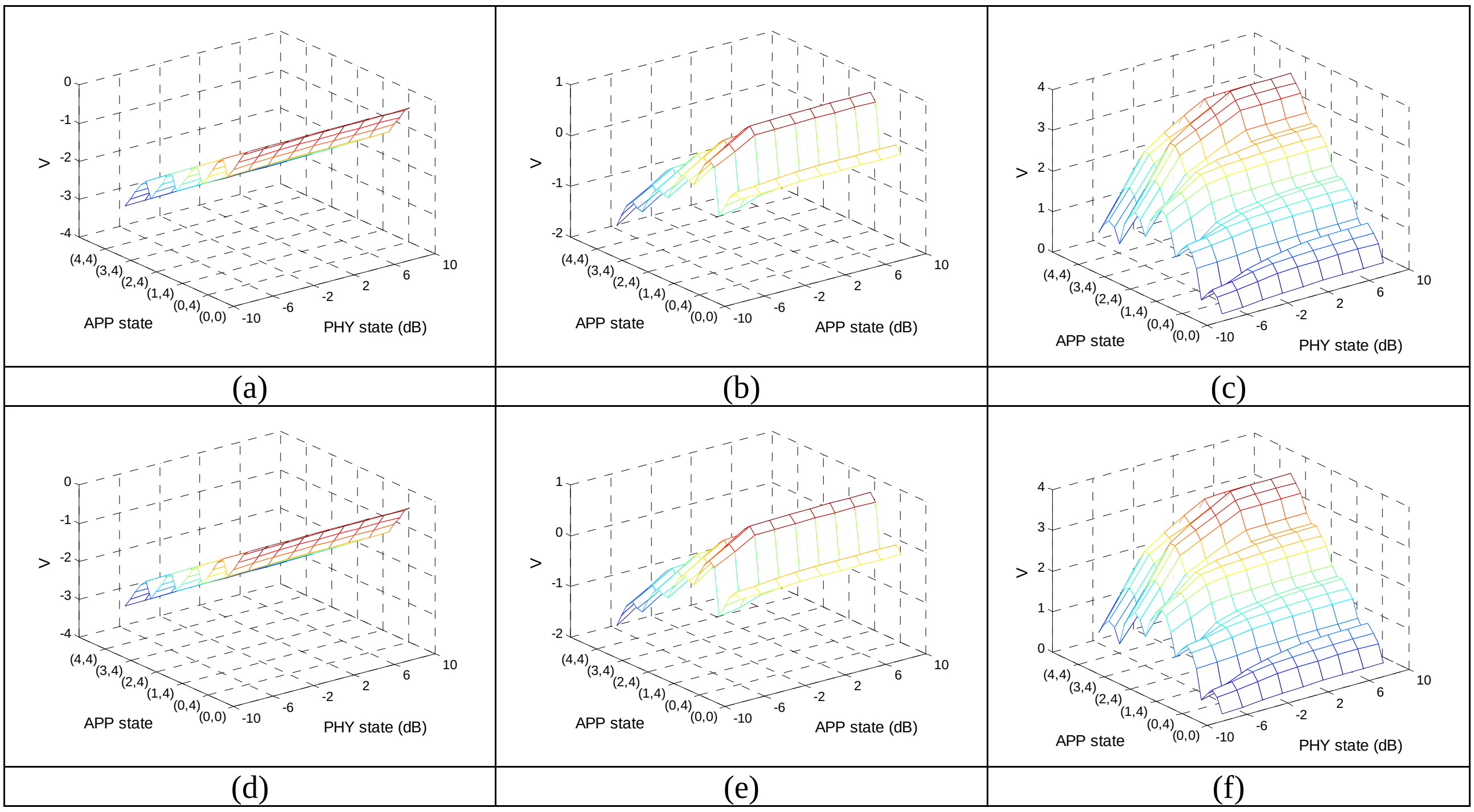

Figure 7. State-value functions resulted from the centralized value iteration and proposed layered value iteration. (a) ~ (c) the value functions of the centralized value iteration when $s_2$ =0.1, 0.5, 1, respectively; (d) ~ (f) the value functions of the layered value iteration when $s_2$ =0.1, 0.5, 1, respectively.

### F. *Myopic versus foresighted optimization*

In this simulation, we use the same parameters as in Section VII.E. We compare the performance of the myopic cross-layer optimization (i.e. $\gamma = 0$) versus our proposed foresighted cross-layer optimization. We first run the value iteration to solve the cross-layer optimization off-line and apply the optimal policy on-line. Figure 8 shows the average reward per stage for both the myopic policy and foresighted policy. The average reward obtained by the foresighted policy is 0.3115 while the average reward by the myopic policy is only 0.0132. Note that this reward value is computed based on the utility function given in VII.D and thus, other types of utility functions may have different values. The simulation results demonstrate that the foresighted policy can achieve much better performance (approximately 24 times better in this simulation) than the myopic policy.

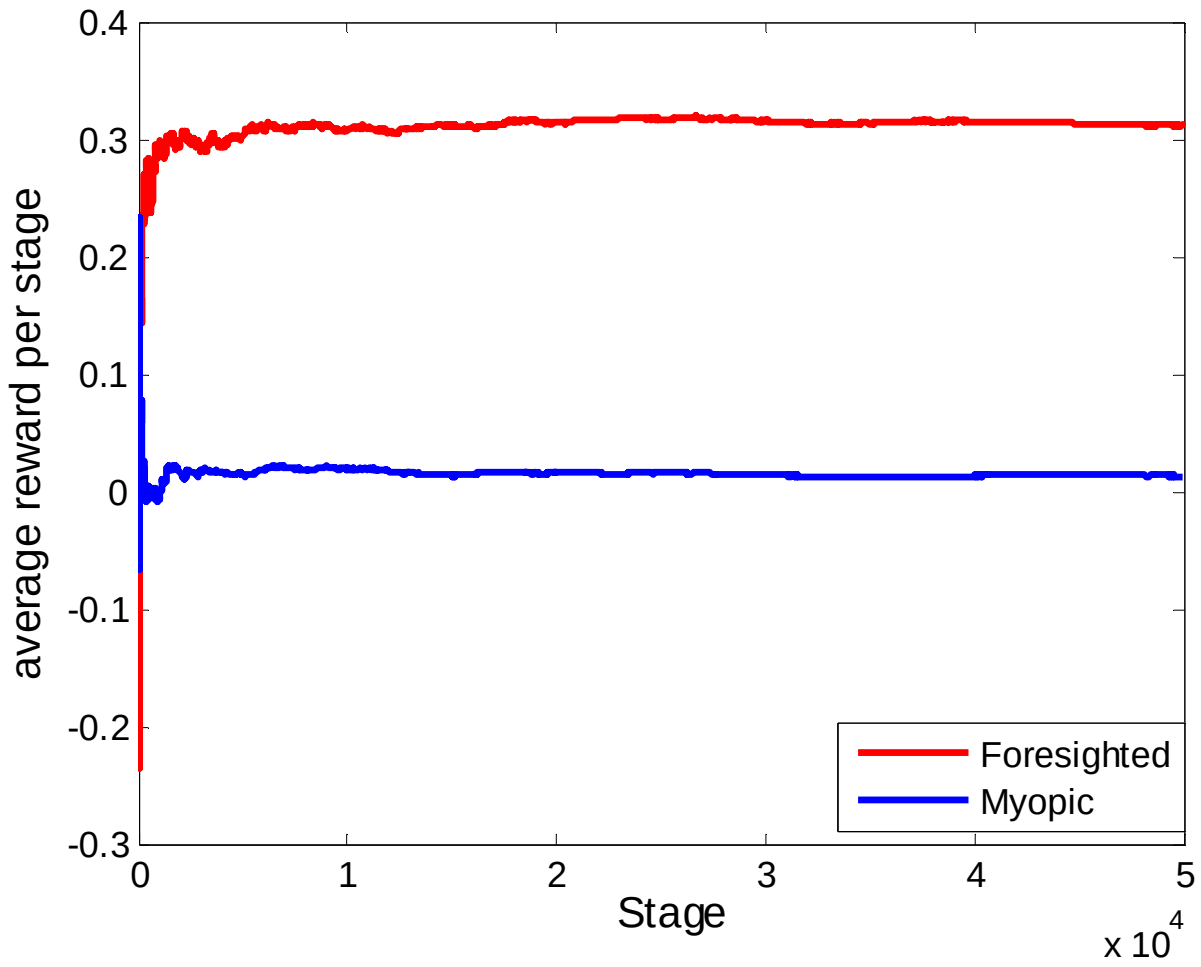

Figure 8.Average reward per state for myopic cross-layer optimization and foresighted cross-layer optimization

### *G. Simplification of layered MDP framework*

In this section, we verify the performance of the simplified version of the layered MDP framework given in Section V.E. The parameters for each layer are the same as in Table 3. In simplification 1, we assume that layer 3 (i.e. the APP layer) models the QoS $Z_{L-1}$ as a uniform distribution among the entire optimal frontier reported by the lower layers. Layer 3 then solves the value iteration shown in Eq. (14). The optimal policy corresponding to this simplified cross-layer optimization is implemented on-line. In simplification 2, we assume that layer 3 does not deploy the external action, i.e. the average packet arrive is constant (equals 1). Then all the layers cooperatively perform the sub-value iterations to find the optimal policy corresponding to this simplification. Then the policy is implemented on-line. The average rewards per stage for both simplifications are shown in Figure 9. To compare performance loss for both simplifications, we also depict the average reward per stage for the layered MDP without simplification. From Figure 9, we note that both simplifications result in suboptimal performance. In this simulation, the reward loss $\Delta R$ due to simplification 1 is around 0.3525, and that due to simplification 2 is around 0.0646. From these results, we can conclude that different simplifications (corresponding to the different amount of information available) have various impacts on the performance gained by the WSTA.

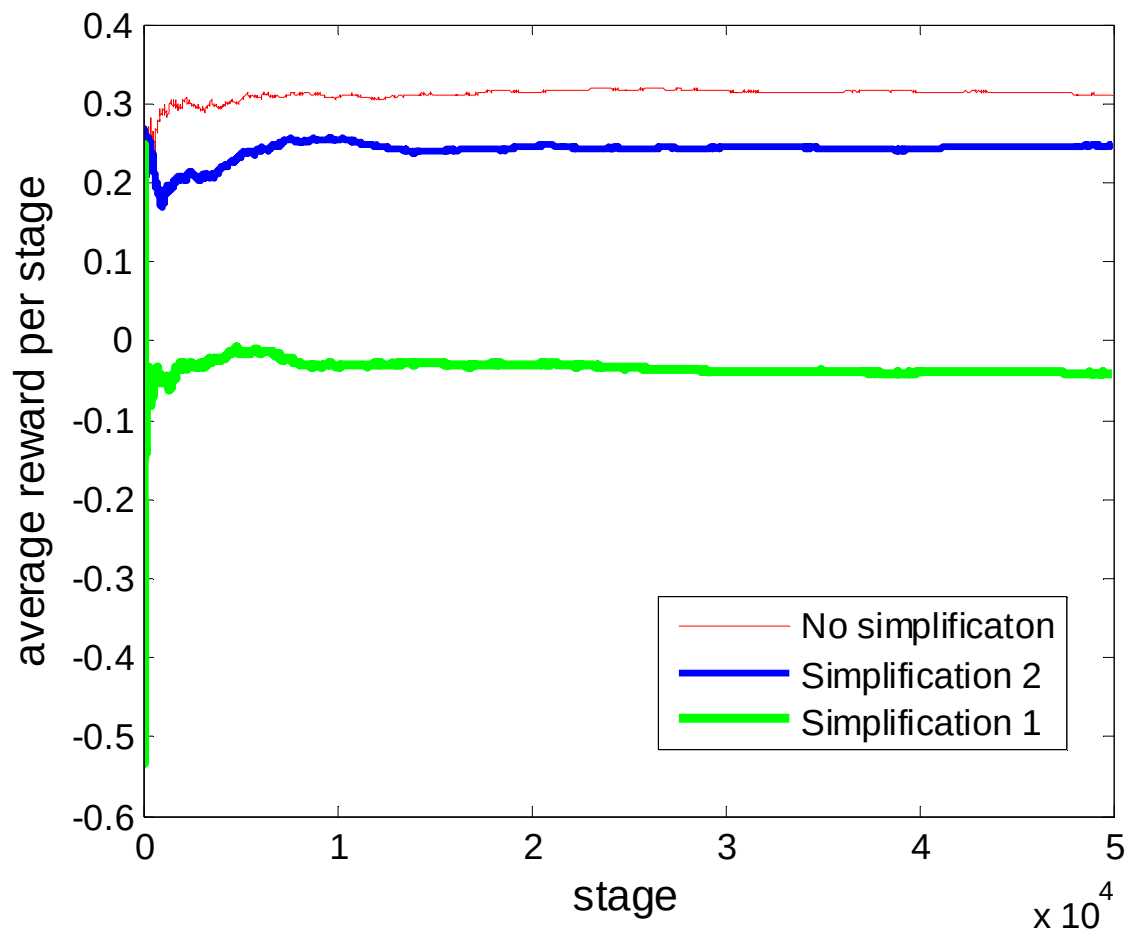

Figure 9. Average reward per stage for the foresighted cross-layer optimization and the simplified versions

### *H. On-line adaptation*

In this simulation, we show that our layered learning algorithm adheres to the layered architecture, while performing as well as the centralized learning algorithm.

The simulation setting is the same as in Section VII.E. However, the dynamics models (transition probabilities) at each layer are unknown. When using the centralized learning algorithm, a central entity within the WSTA is assumed to update the state-value function (critic) and policy (actor) and to choose the action to be performed. When using the layered learning algorithm, each layer updates its own state-value function and policy using the information from other layers, and performs its own action autonomously. The learning parameters are $\alpha = 0.5$ , $\beta = 5$ . Figure 10 shows the average reward achieved by the traditional learning (shown in Section VI.B) and layered learning algorithms (shown in Section VI.C). When learning for enough time, both algorithms approach to the average optimal reward obtained by the optimal policy but still have a gap of approximately 0.1. This performance loss is due to the unknown dynamics models. Interestingly, Figure 10 shows that the proposed layered learning outperforms the traditional one. This can be explained as follows: in the layered learning algorithm, each layer updates its own state-value function and tendency based on the exchanged information and thus, it can obtain a more accurate approximation of the optimal actions.

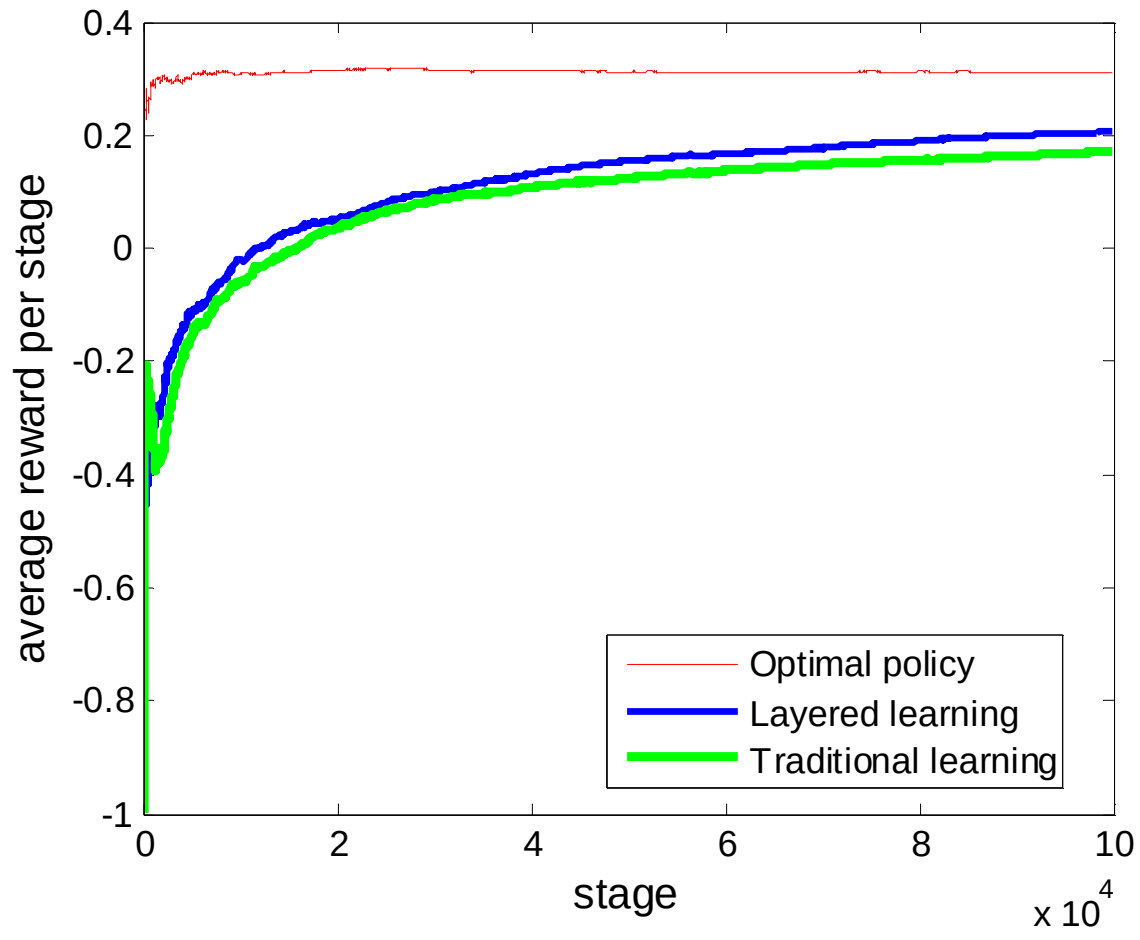

Figure 10. Average reward achieved using both centralized learning and layered learning

## VIII. CONCLUSION

In this paper we formulated the dynamic cross-layer optimization problem as a layered MDP with information exchanges among layers. Within this framework, each layer interacts independently with the environment and experiences different dynamics. The layered MDP is optimally solved using a layered value iteration algorithm. The layered value iteration algorithm allows each layer to perform its own sub-value iteration in order to find the optimal actions in an autonomous manner, given the information exchanges with other layers. Each layer is not required to know the protocols and algorithms implemented at other layers, thereby complying with the current layered network architecture and allowing network designers to build scalable, flexible and upgradable protocols and algorithms. By further examining the layered MDP framework, we show that many existing cross-layer optimization solutions represent simplified versions of our framework. These simplifications lead to suboptimal solutions. The layered MDP framework also allows each layer to autonomously learn its environment dynamics on-line. Our results show that the layered learning algorithm outperforms the traditional learning algorithm for cross-layer optimization.

## Appendix I

Proof: In the layered MDP framework, the layers cooperatively compute the $V_n^*(s)$ by solving the value iteration given in Eq. (9). Based on the reward functions given in Eqs. (3)-(5) and the transition probability in Eq. (2), the value iteration in Eq. (9) can be rewritten as

$$\begin{aligned} V_n^*(\boldsymbol{s}) &= \max_{\boldsymbol{\xi}\in\prod_{l=1}^{L}\Delta_l^a\times\Delta_l^b}\left[R(\boldsymbol{s},\boldsymbol{\xi}) + \gamma\sum_{\boldsymbol{s}'\in\boldsymbol{\mathcal{S}}} p(\boldsymbol{s}'\mid\boldsymbol{s},\boldsymbol{\xi})V_{n-1}^*(\boldsymbol{s}')\right] \\ &= \max_{\boldsymbol{\xi}\in\prod_{l=1}^{L}\Delta_l^a\times\Delta_l^b}\left[R_{in}(\boldsymbol{s},\boldsymbol{b}) - \sum_{l=1}^{L}\lambda_l c_l(s_l,a_l) + \gamma\sum_{\boldsymbol{s}'\in\boldsymbol{\mathcal{S}}}\prod_{l=1}^{L-1} p(s_l\mid\boldsymbol{s}'_{1\to l-1},s_l,a_l)p(s'_L\mid\boldsymbol{s}'_{1\to L-1},\boldsymbol{s},a_L,\boldsymbol{b})V_{n-1}^*(\boldsymbol{s}')\right] \end{aligned} \tag{33}$$

Given the upward message $\theta_{L-1,L} = \mathcal{Z}_{L-1}$, layer $L$ can generate the optimal frontier $\mathcal{Z}_L$. Then, the iteration in Eq. (33) becomes

$$\begin{aligned} &V_n^*(\boldsymbol{s}) = \\ &\max_{\substack{\boldsymbol{a}\in\prod_{l=1}^{L}\Delta_l^a \\ Z_L\in\mathcal{Z}_L}}\left[R_{in}(s_L,Z_L) - \sum_{l=1}^{L}\lambda_l c_l(s_l,a_l) + \gamma\sum_{\boldsymbol{s}'\in\boldsymbol{\mathcal{S}}}\prod_{l=1}^{L-1} p(s'_l\mid\boldsymbol{s}'_{1\to l-1},s_l,a_l)p(s'_L\mid\boldsymbol{s}'_{1\to L-1},\boldsymbol{s},a_L,Z_L)V_{n-1}^*(\boldsymbol{s}')\right] \\ &= \max_{\substack{\boldsymbol{a}\in\prod_{l=1}^{L}\Delta_l^a \\ Z_L\in\mathcal{Z}_L}}\left\{-\sum_{l=1}^{L-1}\lambda_l c_l(s_l,a_l) + \sum_{\boldsymbol{s}'_{1\to L-1}\in\boldsymbol{\mathcal{S}}_{1\to L-1}}\prod_{l=1}^{L-1} p(s'_l\mid\boldsymbol{s}'_{1\to l-1},s_l,a_l)\left[\begin{array}{l} R_{in}(s_L,Z_L) - \lambda_L c_L(s_L,a_L) + \\ \gamma\sum_{s'_L\in\mathcal{S}_L} p(\boldsymbol{s}'_L\mid\boldsymbol{s}'_{1\to L-1},\boldsymbol{s}_L,a_L,Z_L)V_{n-1}^*(\boldsymbol{s}')\end{array}\right]\right\} \\ &= \max_{a_{1\to L-1}\in\prod_{l=1}^{L-1}\Delta_l^a}\left\{-\sum_{l=1}^{L-1}\lambda_l c_l(s_l,a_l) + \sum_{\boldsymbol{s}'_{1\to L-1}\in\boldsymbol{\mathcal{S}}_{1\to L-1}}\prod_{l=1}^{L-1} p(s'_l\mid\boldsymbol{s}'_{1\to l-1},s_l,a_l)\underbrace{\max_{\substack{a_L\in\Delta_L^a, \\ Z_L\in\mathcal{Z}_L}}\left[\begin{array}{l} R_{in}(s_L,Z_L) - \lambda_L c_L(s_L,a_L) \\ +\gamma\sum_{s'_L\in\mathcal{S}_L} p(\boldsymbol{s}'_L\mid\boldsymbol{s}'_{1\to L-1},\boldsymbol{s}_L,a_L,Z_L)V_{n-1}^*(\boldsymbol{s}')\end{array}\right]}_{\text{value iteration of layer } L}\right\} \end{aligned} \tag{34}$$

The transition from the first line to second line in Eq. (34) is because $R_{in}(s_L,Z_L) - \lambda_L c_L(s_L,a_L)$ is independent of the states $s'_{1\to L-1}$ and $\sum_{\boldsymbol{s}'_{1\to L-1}\in\boldsymbol{\mathcal{S}}_{1\to L-1}}\prod_{l=1}^{L-1} p(s'_l\mid\boldsymbol{s}'_{1\to l-1},s_l,a_l) = 1$ and $p(\boldsymbol{s}'_L\mid\boldsymbol{s}'_{1\to L-1},\boldsymbol{s}_L,a_L,Z_L) = p(\boldsymbol{s}'_L\mid\boldsymbol{s}'_{1\to L-1},\boldsymbol{s},a_L,\boldsymbol{b})$ given $Z_L$.

Since layer $L$ is not allowed to know the actions and dynamics at lower layers, layer $L$ performs the value iteration for each state $\boldsymbol{s}'_{1\to L-1}$ as follows

$$V_{n,L-1}^*(\boldsymbol{s}'_{1\to L-1}) = \max_{\substack{a_L\in\Delta_L^a, \\ Z_L\in\mathcal{Z}_L}}\left[R_{in}(s_L,Z_L) - \lambda_L c_L(s_L,a_L) + \gamma\sum_{s'_L\in\mathcal{S}_L} p(\boldsymbol{s}'_L\mid\boldsymbol{s}'_{1\to L-1},\boldsymbol{s}_L,a_L,Z_L)V_{n-1}^*(\boldsymbol{s}')\right] \tag{35}$$

We can interpret $V_{n,L-1}^*(\boldsymbol{s}'_{1\to L-1})$ as state-value function of state $\boldsymbol{s}'_{1\to L-1}$ seen at layer $L-1$ at iteration $n$. After performing the value iteration, layer $L$ sends the message $\theta_{L,L-1} = \left\{V_{n,L-1}^*(\boldsymbol{s}'_{1\to L-1})\right\}$ to layer $L-1$. The message $\left\{V_{n,L-1}^*(\boldsymbol{s}'_{1\to L-1})\right\}$ represents the expected future reward for the layers $\{1,...,L-1\}$.

After layer $L$ solves its own sub-value iteration, the value iteration in Eq. (34) becomes

$$
\begin{aligned}
V_n^*(\boldsymbol{s}) &= \max_{a_{1\to L-1}\in\prod_{l=1}^{L-1}\Delta_l^a}\left\{-\sum_{l=1}^{L-1}\lambda_l c_l(s_l,a_l)+\sum_{\boldsymbol{s}'_{1\to L-1}\in\boldsymbol{\mathcal{S}}_{1\to L-1}}\prod_{l=1}^{L-1}p\left(s'_l\mid \boldsymbol{s}'_{1\to l-1},s_l,a_l\right)V_{n,L-1}^*\left(\boldsymbol{s}'_{1\to L-1}\right)\right\}\\
&= \max_{\boldsymbol{a}_{1\to L-2}\in\prod_{l=1}^{L-2}\Delta_l^a}\left\{-\sum_{l=1}^{L-2}\lambda_l c_l(s_l,a_l)+\right.\\
&\left.\sum_{\boldsymbol{s}'_{1\to L-2}\in\boldsymbol{\mathcal{S}}_{1\to L-2}}\prod_{l=1}^{L-2}p\left(s'_l\mid \boldsymbol{s}'_{1\to l-1},s_l,a_l\right)\underbrace{\max_{a_{L-1}\in\Delta_{L-1}^a}\left[\begin{array}{l}-\lambda_{L-1}c_{L-1}(s_{L-1},a_{L-1})\\+\sum_{\boldsymbol{s}'_{L-1}\in\mathcal{S}_{L-1}}p\left(s'_{L-1}\mid \boldsymbol{s}'_{1\to L-2},s_{L-1},a_{L-1}\right)V_{n,L-1}^*\left(\boldsymbol{s}'_{1\to L-1}\right)\end{array}\right]}_{\text{value iteration of layer } L\text{-1}}\right\}
\end{aligned}
\tag{36}
$$

The transition from the first line to second line in Eq. (36) has the similar interpretation as in Eq. (34). Without knowing the actions and transition probability at lower layers, layer $L-1$ performs the sub-value iteration for each state $\boldsymbol{s}'_{1\to L-2}$ as follows

$$
V_{n-1}^*\left(\boldsymbol{s}'_{1\to L\text{-}2}\right)=\max_{a_{L-1}\in\Delta_{L-1}^a}\left[-\lambda_{L-1}c_{L-1}(s_{L-1},a_{L-1})+\sum_{\boldsymbol{s}'_{L-1}\in\mathcal{S}_{L\text{-}1}}p\left(s'_{L-1}\mid \boldsymbol{s}'_{1\to L-2},s_{L-1},a_{L-1}\right)V_{n,L-1}^*\left(\boldsymbol{s}'_{1\to L-1}\right)\right] \tag{37}
$$

Similarly, for each state $\boldsymbol{s}'_{1\to l-1}$, layer $l$ performs the sub-value iteration as follows

$$
V_{n,l\text{-}1}^*\left(\boldsymbol{s}'_{1\to l\text{-}1}\right)=\max_{a_l\in\Delta_l^a}\left[-\lambda_l c_l(s_l,a_l)+\sum_{s'_l\in\mathcal{S}_l}p\left(s'_l\mid \boldsymbol{s}'_{1\to l-1},s_l,a_l\right)V_{n,l}^*\left(\boldsymbol{s}'_{1\to l}\right)\right] \tag{38}
$$

We can interpret $V_{n,l\text{-}1}^*\left(\boldsymbol{s}'_{1\to l\text{-}1}\right)$ as state-value function of state $\boldsymbol{s}'_{1\to l\text{-}1}$ seen at layer $l-1$ at iteration $n$. The message exchanged from layer $l$ to layer $l-1$ is $\theta_{l,l-1}=\left\{V_{n,l}^*\left(\boldsymbol{s}'_{1\to l-1}\right)\right\}$.

At layer 1, the value iteration is

$$
V_n^*(\boldsymbol{s})=\max_{a_1\in\Delta_l^a}\left[-\lambda_1 c_1(s_1,a_1)+\sum_{s'_1\in\mathcal{S}_1}p\left(\boldsymbol{s}'_1\mid \boldsymbol{s}_1,a_1\right)V_{n,1}^*\left(\boldsymbol{s}'_1\right)\right] \tag{39}
$$

By re-denote the state-value function $V_n^*(\boldsymbol{s})$ as $V_{n,L}^*(\boldsymbol{s}_{1\to L})$, we can obtain the sub-value iteration at each layer as shown in Table 1.

# Appendix II

In this appendix, we show that the internal reward function satisfies the preservation of the internal reward function property.

Recall that $v_3^k=\left\lfloor\delta\left(1-\varepsilon_3^k\right)/t_3^k\right\rfloor$. We note that $v_{APP}^k$ is a non-increasing function of both $\varepsilon_3^k$ and $t_3^k$. Using Definition 4, $Z_3^k\overset{d.}{\le}\tilde{Z}_3^k$ means that $\varepsilon_3^k\le\tilde{\varepsilon}_3^k$ and $t_3^k\le\tilde{t}_3^k$ and $f_3^k\le\tilde{f}_3^k$. Hence, we have $v_3^k\left(Z_3^k\right)\ge v_3^k\left(\tilde{Z}_3^k\right)$.

Given $s_{3,1}^k$, the function $\min\left\{s_{3,1}^k-v_3^k\left(Z_3^k\right),0\right\}$ is a non-increasing function of $v_3^k$ and hence, we have $\min\left\{s_{3,1}^k-v_3^k\left(Z_3^k\right),0\right\}\le\min\left\{s_{3,1}^k-v_3^k\left(\tilde{Z}_3^k\right),0\right\}$.

Hence, we have

$$v_3^k\left(Z_3^k\right)-\lambda_g\min\left\{s_{3,1}^k-v_3^k\left(Z_3^k\right),0\right\}-f_3^k\geq v_3^k\left(\tilde{Z}_3^k\right)-\lambda_g\min\left\{s_{3,1}^k-v_3^k\left(\tilde{Z}_3^k\right),0\right\}-\tilde{f}_3^k \quad (40)$$

That is, $R_{in}\left(s_3^k,Z_3^k\right)\geq R_{in}\left(s_3^k,\tilde{Z}_3^k\right)$.